\documentclass[onecolumn,showpacs,aps,prl,floatfix,citeautoscript
,cite,nofootinbib]{revtex4-2}
\usepackage{graphicx}
\usepackage{epstopdf}
\usepackage{color}
\usepackage{amsmath}
\usepackage{ulem}
\usepackage{hyperref}

\begin{document}

\title{Beyond the orbitally–resolved magnetic exchange in CrI$_{3}$ and NiI$_{2}$}

\author{D. \v{S}abani}
\email{denis.sabani@uantwerpen.be}
\affiliation{Department of Physics \& NANOlab Center of Excellence, University of Antwerp, Groenenborgerlaan 171, B-2020 Antwerp, Belgium}

\author{C. Bacaksiz}
\affiliation{Department of Physics \& NANOlab Center of Excellence, University of Antwerp, Groenenborgerlaan 171, B-2020 Antwerp, Belgium}

\author{M. V. Milo\v{s}evi\'c}
\email{milorad.milosevic@uantwerpen.be}
\affiliation{Department of Physics \& NANOlab Center of Excellence, University of Antwerp, Groenenborgerlaan 171, B-2020 Antwerp, Belgium}

\begin{abstract}
The pertinent need for microscopic understanding of magnetic exchange motivated us to go beyond the existing theories and develop a systematic method to quantify \textit{all} possible mechanisms that contribute to magnetic exchange for an arbitrary pair of atoms in a given material. 
We apply it to the archetypal 2D magnetic monolayers CrI$_{3}$ and NiI$_{2}$, to reveal the previously underrated $d_{x^{2}-y^{2}}$,$d_{x^{2}-y^{2}}$ contribution as either the leading or the second largest contribution to the total magnetic exchange. We proceed to explore the microscopic mechanisms behind all the non–zero orbital contributions in both CrI$_{3}$ and NiI$_{2}$, and generalize the findings to other magnetic monolayers dominated by $d^{8}$ and $d^{3}$ electronic configurations of the magnetic atoms.
\end{abstract}

\date{\today}
\pacs{Valid PACS appear here}
\maketitle

\section{Introduction} 
Since the first unambiguous experimental demonstration of long–range magnetism in atomically thin materials in 2017~\cite{CrI32017,CrGeTe2017}, disobeying the Mermin–Wagner theorem~\cite{MerminWagner}, the class of magnetic two–dimensional materials (M2DMs) has attracted enormous interest from the research community~\cite{CrGeTe2017_Xing,CrGeTe2017_Chen,VSe2,Efield_switch,MnSex,CrCl3-xBr3,Giant_tunnel,2DvdW_etunnel,onemilion,Deng2018_FeGeTe}, for their fundamental importance, susceptibility to practical manipulations such as gating~\cite{CrGeTe2017_Xing,CrGeTe2017_Chen,Efield_switch,Deng2018_FeGeTe}, stacking~\cite{sivadas2018,song2019,thiel2019,li2019,song2021}, and strain~\cite{dai2019,li2019,hsu2020,cenker2022}, the ability to retain~\cite{VSe2,ohara2018} or 
attain~\cite{Deng2018_FeGeTe} magnetism beyond room temperature, and their suitability for van der Waals and moir\'e heterostructuring ~\cite{song2021,jin2022,xie2023} toward multifunctional hybrid materials and devices. Moreover, multiferroic behavior has been experimentally observed in monolayer and few–layer NiI$_{2}$~\cite{song2022,amini2024}, where the non–collinear magnetic order breaks inversion symmetry and directly couples to the non–zero ferroelectric polarization.

This persistently booming experimental research on M2DMs was accompanied by numerous timely and important theoretical studies~\cite{Lado2017,liu2018,webster2018,Kitaev2018,sivadas2018,torelli2018,woojang2019,kim2019,jiang2019,soriano2019,zhang2019,Kashin2020,cxu2020,kvashnin2020,pizzochero2020,bacaksiz2021,ni2021,cxu2022,fumega2022,edstrom2022,riedl2022,bo2023,bo2023b,wang2023,bo2024,yorulmaz2024,liu2024}. Obviously, such hand–in–hand progress remains crucial for detailed understanding and proper theoretical description of the magnetic interactions, their consequences, and their sources in the M2DMs. The current workflow in the literature on this topic includes calculations of atomic and electronic structure of a material, and subsequent extraction of magnetic exchange parameters that constitute the minimal Heisenberg Hamiltonian ($H_{H} = \sum_{i < j} \vec{S}_{i}[J_{ij}]_{3\times 3}\vec{S}_{j}$, $[J_{ij}]$ being the interatomic magnetic exchange matrix) using the first–principles methods. The obtained magnetic exchange parameters are then used to explain different experimental observables, such as magnetic order in the system of interest~\cite{Lado2017,liu2018,webster2018,Kitaev2018,Kashin2020,pizzochero2020,bacaksiz2021,sivadas2018,woojang2019,jiang2019,soriano2019,fumega2022}, its critical temperature~\cite{torelli2018,kim2019,bacaksiz2021}, etc. While these consequences of magnetic exchange are well understood, the exact quantitative determination of magnetic exchange between a pair of atoms for a specific material remains challenging. Riedl \textit{et al.}~\cite{riedl2022} reported the worryingly strong dependence of magnetic exchange parameters on the computational details. For example, they found that in case of calculations for NiI$_{2}$, changing the type of energy mapping and the value of on–site Coulomb repulsion (U parameter) results in a broad range of values for ferromagnetic (FM) first–nearest–neighbor (1NN) magnetic exchange ($-1.2$~meV to $-8.0$~meV), and for antiferromagnetic (AFM) third–nearest–neighbor (3NN) magnetic exchange ($+1.8$~meV to $+5.8$~meV).
It therefore became clear that, instead of fiddling with the computational setup, one should rather focus on understanding the origin of the obtained value for the desired magnetic exchange parameter. Quite the contrary, in the current state–of–the–art, the origin of the obtained value for magnetic exchange very often remains entirely unaddressed, or when discussed, it is mostly illustrated using a schematic representation of the direct exchange and/or mediated superexchange processes\footnote{For clarification about still present confusion in the literature about direct/super and potential/kinetic exchange we refer to Ref.~\onlinecite{anderson2}.}, without the exact calculation to support the premise~\cite{Kitaev2018,soriano2019,zhang2019,riedl2022,wang2023}. The minority of the works to date went a step further, to exactly calculate the magnetic interaction between the $d$ orbitals of magnetic atoms~\cite{woojang2019,Kashin2020,kvashnin2020}, but even in those analyses, discussion is guided towards the suggested schematic diagrams of those particular $d$–orbitals interacting directly, or via mediating $p$–orbital(s) of non–magnetic atoms, omitting the other possible orbital contributions.

\begin{figure}[t]
\centering
\includegraphics[width=\columnwidth]{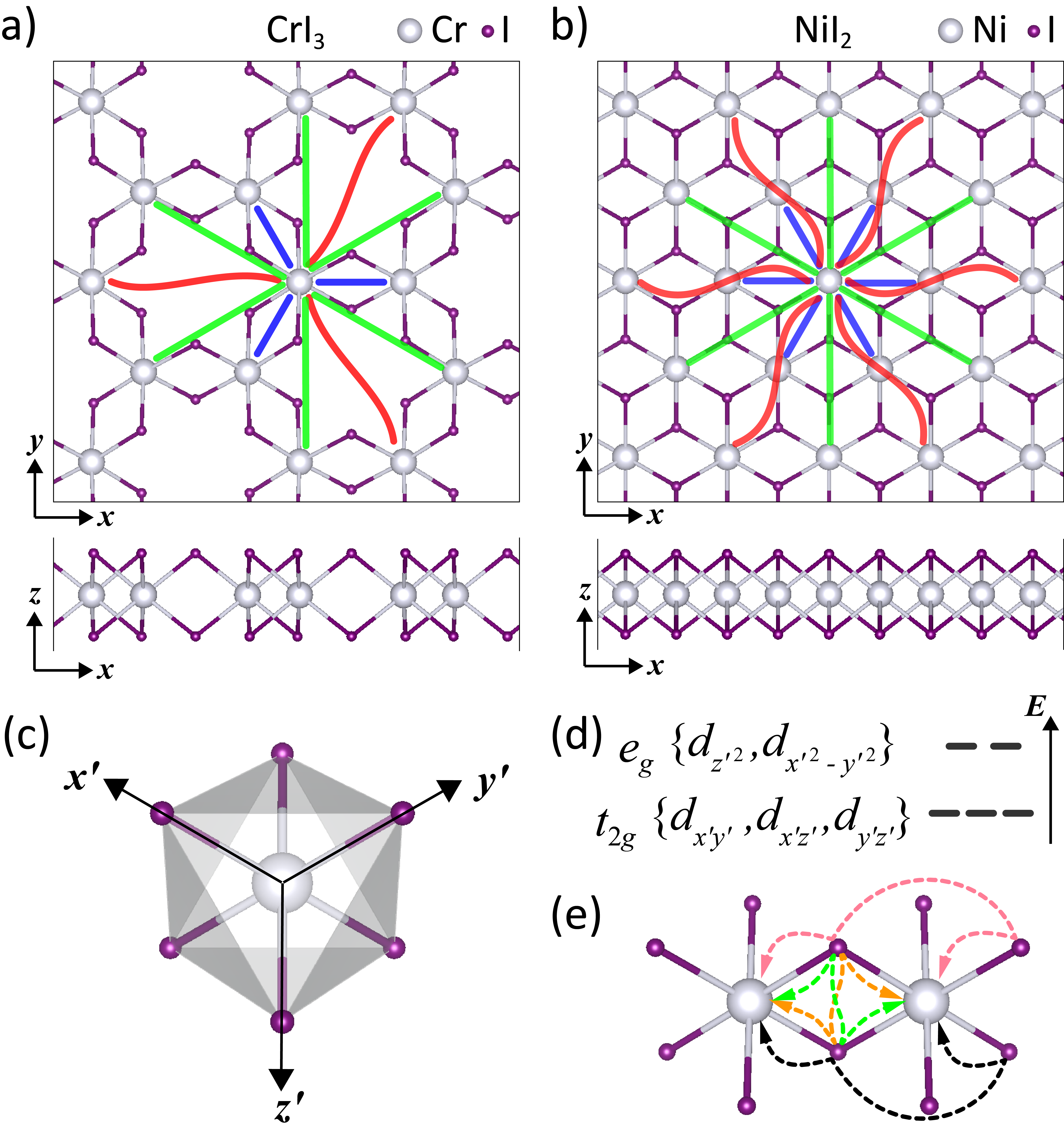}
\caption{\textbf{Global and local structural properties.} Top and side view of archetypal magnetic 2D monolayers (a) CrI$_{3}$ and (b) NiI$_{2}$, with global in–plane ($x,y$) and out–of–plane ($z$) coordinates indicated. Blue, green and red lines connect 1NN, 2NN and 3NN pairs of TM atoms. (c) Local octahedral surrounding of a TM atom in both materials, with local octahedral coordinates indicated as $x'$, $y'$ and $z'$, and (d) corresponding energy level splitting of $d$ orbitals of the TM atom. (e) Illustration of four super–superexchange paths.}
\label{fig1}
\end{figure}
Therefore, to remedy this unsatisfactory picture, we present here a method that objectively quantifies the contributions of \textit{all} microscopic processes behind \textit{each} orbitally–resolved contribution in the magnetic exchange interaction between \textit{any} two transition–metal (TM) atoms in a given material. We focus on the much studied CrI$_{3}$ (Fig.~\ref{fig1}a) and NiI$_{2}$ (Fig.~\ref{fig1}b) monolayers, where TM atoms have $d^{3}$ and $d^{8}$ electronic occupation, respectively. Both materials host a local octahedral environment ($x'$, $y'$, and $z'$ in Fig.~\ref{fig1}c) and their $d$ levels are split into $t_{2g}$ and $e_{g}$ many–folds (Fig.~\ref{fig1}d). We reveal the significance of the $d_{x'^{2}-y'^{2}}$,$d_{x'^{2}-y'^{2}}$ contribution and its origin as electron hopping mediated by two non–magnetic atoms, a mechanism we refer to as ``super–superexchange'' (see Fig.~\ref{fig1}e). Last but not least, we discuss the influence of the specific electronic configuration of the constituent TM atoms, on the resulting magnetic interactions in a given M2DM.

\section{Theoretical approach} In order to quantify magnetic exchange between TM atoms in the system, and its orbitally–resolved constituents, we use Lichtenstein–Katsnelson–Antropov–Gubanov (LKAG)~\cite{LKAG1987} formalism based on the second–order energy mapping between $H_{H}$ and the density functional theory (DFT) Hamiltonian in a localized basis, as implemented in TB2J~\cite{TB2J2021}. All DFT calculations were performed using the Vienna Ab initio Simulation Package (VASP)~\cite{vasp1,vasp2,vasp3} and Wannier functions (WFs)~\cite{wannier90} were chosen for the localized basis set. For a thorough description of all computational details, we refer to the Supplemental Material (SM)\cite{SM}, including its references\cite{dudarevU, TB2J2021, moriya, Kitaev2018, amoroso2020}.

To quantify the influence of different microscopic mechanisms on magnetic exchange, we established a \textit{successive hopping inclusion method} (SHIM). Orbitally–resolved magnetic exchange parameters are determined by electron hopping parameters between different orbitals and orbitals' on–site energies. Magnetic exchange between any two $d$–orbitals on two magnetic atoms will be non–zero if and only if there exists at least one hopping path connecting those two orbitals through the unbroken chain of non–zero hopping parameters (cf. Fig.~\ref{fig1}e). 

To extract the significance to magnetic exchange of each orbitally–resolved closed hopping path, we gradually reduce the threshold of hoppings considered ($|t| > \tau$) from above the largest hopping value (no hopping allowed) to 0 (all hoppings included). As $\tau$ is decreased, hopping paths are successively activated, each causing a detectable change in the magnetic exchange, enabling us to pinpoint contributions of specific microscopic processes in the orbitally–resolved picture. 
For all the details of this method, we refer to the SM \cite{SM}.

\begin{figure}[t]
\centering
\includegraphics[width=\columnwidth]{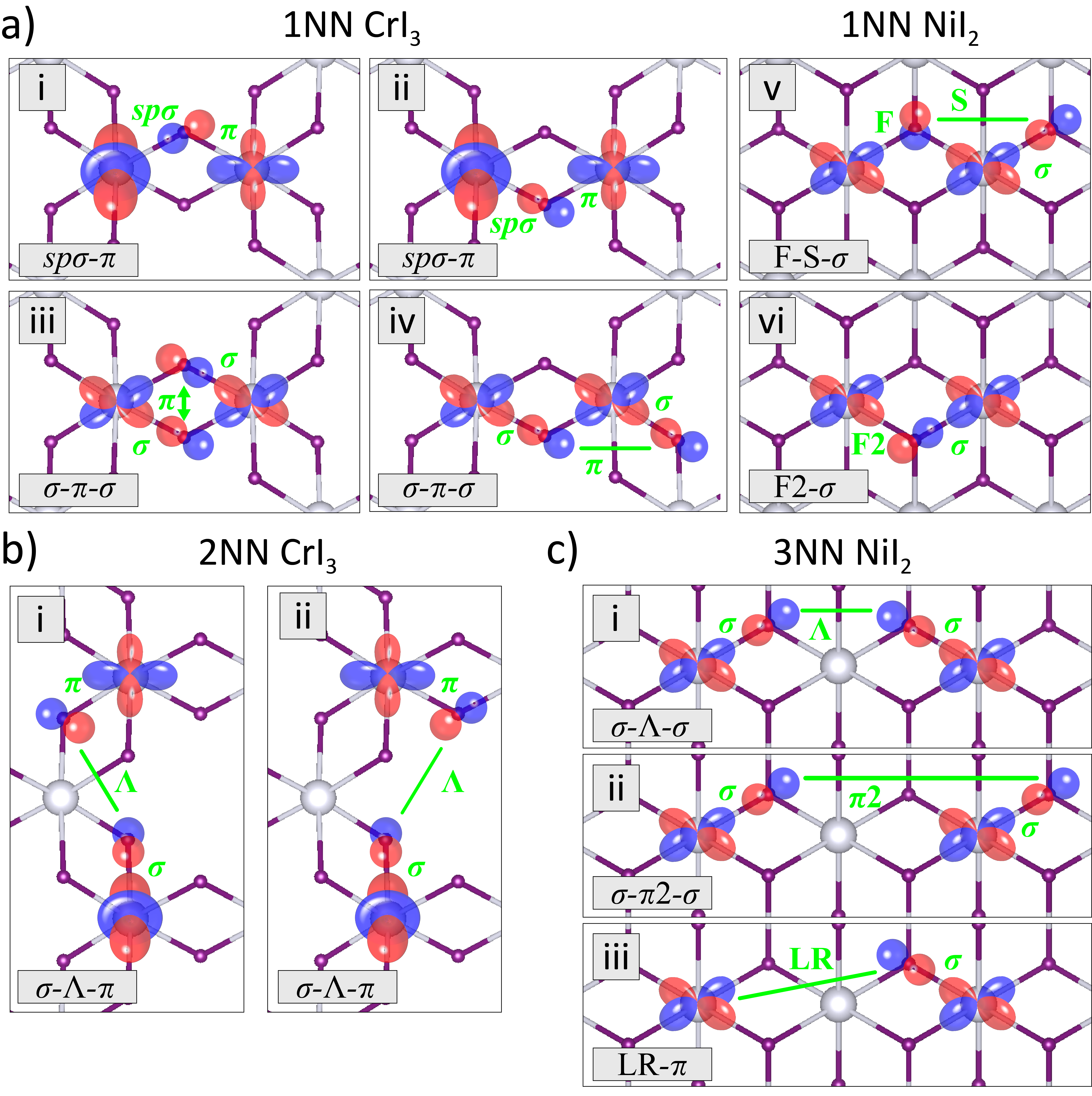}
\caption{\textbf{Superexchange versus super–superexchange paths.} 
(a) 1NN magnetic super–superexchange paths in CrI$_{3}$ and NiI$_{2}$: (i) and (ii) “$sp\sigma$–$\pi$”; (iii) and (iv) “$\sigma$–$\pi$–$\sigma$”; (v) “F–S–$\sigma$”; (vi) “F'–$\sigma$”. 
(b) 2NN magnetic super–superexchange paths in CrI$_{3}$: (i) and (ii) “$\sigma$–$\Lambda$–$\pi$”.
(c) 3NN magnetic super–superexchange paths in NiI$_{2}$: (i) “$\sigma$–$\Lambda$–$\sigma$”; (ii) “$\sigma$–$\pi$'–$\sigma$”; (iii) “LR–$\sigma$”.}
\label{fig_closed_paths}
\end{figure}
\section{Hopping parameters and paths} 
Let us first identify and label different types of hopping parameters that constitute the paths (see Fig.~\ref{fig_closed_paths}) relevant for the dominant 1NN and 2NN magnetic exchange in CrI$_{3}$, and 1NN and 3NN magnetic exchange in NiI$_{2}$. The hopping that connects two orbitals whose lobes are pointing directly towards each other is labeled $\sigma$, borrowing the terminology from chemistry. Analogously, the hopping between two orbitals whose axes are parallel to each other is labeled $\pi$, following the definition of a $\pi$ bond. The hopping where the lobe of a $p$ orbital points towards the torus of $d_{z'^{{2}}}$ is labeled as $sp\sigma$, because the torus–lobe overlap resembles the sphere–lobe $sp$ overlap in the case of e.g. H–C bond in organic compounds. The labels $F$ and $F'$ are assigned to hopping parameters between $d$ and $p$ orbitals which would be forbidden in ideal octahedral surrounding, but are non–zero due to the slight distortions present. The hopping parameters between two $p$ orbitals whose axes are misaligned are labeled $S$ if skewed, and $\Lambda$ if orthogonal to each other. The hopping of $\pi$ type, however between atoms distant apart, is labeled $\pi'$. Finally, the long–range hopping between $d_{x^{2}-y^{2}}$ orbital and $p$ orbital on a distant iodine atom is labeled $LR$.

\section{The case of CrI$_{3}$} In what follows, we present the results for the isotropic part of the magnetic exchange, responsible for FM/AFM alignment of magnetic moments, and dissect all non–zero orbitally–resolved contributions w.r.t. the relevant closed hopping paths. The anisotropic part is presented in the SM \cite{SM}. 
We commence with monolayer CrI$_{3}$ where only 1NN and 2NN magnetic exchange are non–zero and FM, namely $J_{1NN} = -3.72$~meV and $J_{2NN} = -2.10$~meV. In our analysis, the 1NN pair considered will be the one coinciding with the global $x$ direction and the 2NN pair the one coinciding with the global $y$ direction (cf. Fig. \ref{fig1}a). 

\begin{figure}[t]
\centering
\includegraphics[width=\columnwidth]{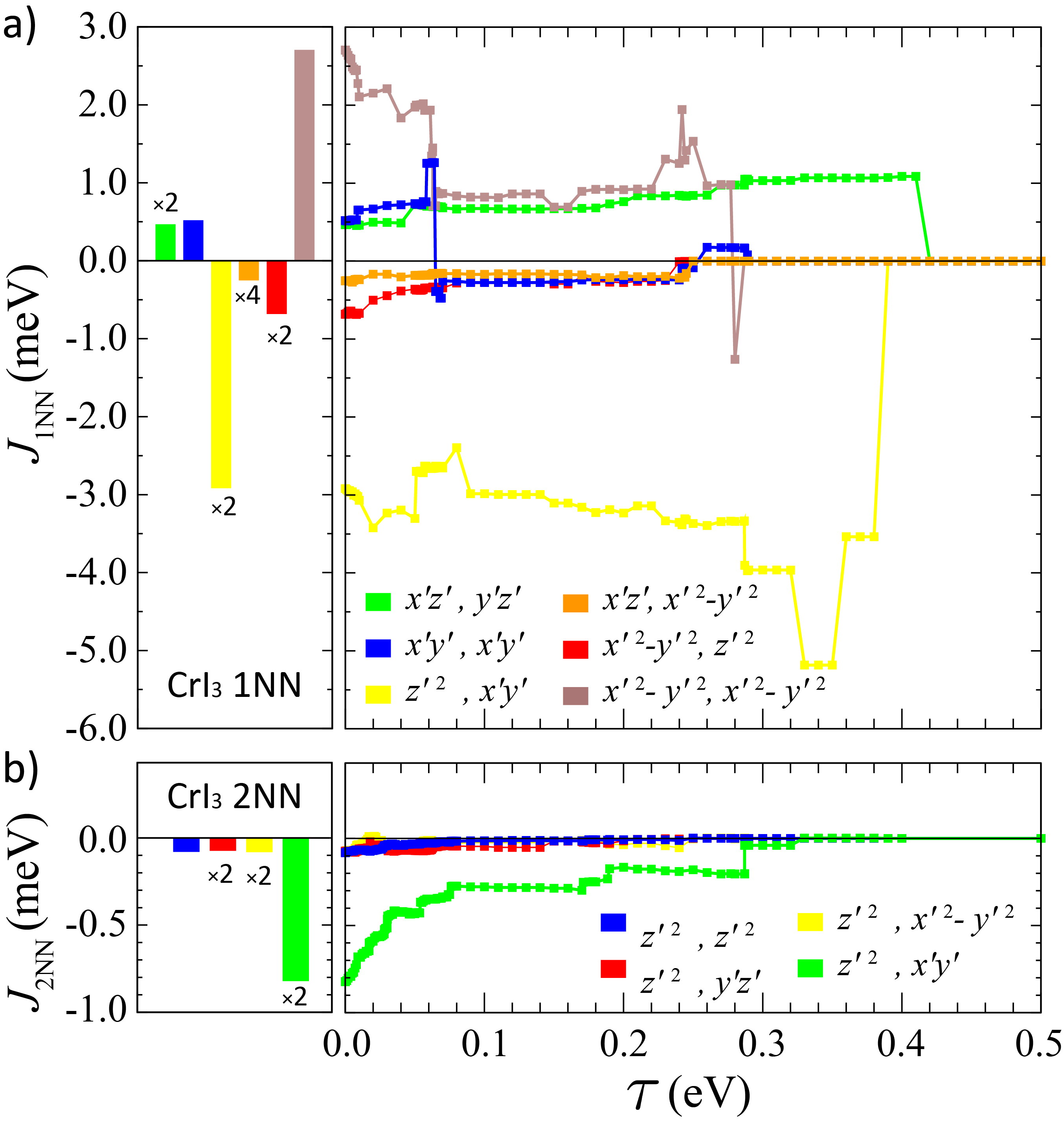}
\caption{\textbf{Origin of $J_{1NN}$ and $J_{2NN}$ in CrI$_{3}$.} (a) Orbitally–resolved contributions to the 1NN magnetic exchange (bars) and their hopping–dependent origin (lines), in monolayer CrI$_{3}$. (b) Same as (a), for the 2NN magnetic exchange.}
\label{fig_CrI3}
\end{figure}

In case of $J_{1NN}$, out of 9 possible $t_{2g},t_{2g}$ contributions, only 3 are non–zero and all are AFM: $d_{x'z'},d_{y'z'}$, $d_{y'z'},d_{x'z'}$ and $d_{x'y'},d_{x'y'}$ (green and blue bars in Fig.~\ref{fig_CrI3}a). Further, the 12 possible $t_{2g},e_{g}$ and $e_{g},t_{2g}$ contributions reduce to 6 non–zero ones, all being FM: $d_{x'y'},d_{z'^{2}}$, $d_{z'^{2}},d_{x'y'}$ and $d_{x'z'},d_{x'^{2}-y'^{2}}$, $d_{y'z'},d_{x'^{2}-y'^{2}}$, $d_{x'^{2}-y'^{2}},d_{x'z'}$, $d_{x'^{2}-y'^{2}},d_{y'z'}$ (yellow and orange bars in Fig.~\ref{fig_CrI3}a). Finally, out of 4 possible $e_{g},e_{g}$ contributions 3 are non–zero: the twins $d_{z'^{2}},d_{x'^{2}-y'^{2}}$ and $d_{x'^{2}-y'^{2}},d_{z'^{2}}$ are FM (red bar in Fig.~\ref{fig_CrI3}a) and $d_{x'^{2}-y'^{2}},d_{x'^{2}-y'^{2}}$ is AFM (brown bar in Fig.~\ref{fig_CrI3}a).

The $d_{z'^{2}},d_{x'y'}$ contribution dominates and determines the total FM character of 1NN magnetic exchange in CrI$_{3}$. It originates from the sp$\sigma$–$\pi$ superexchange, mediated by one iodine atom (see Fig.~\ref{fig_closed_paths}a). For $\tau = 0.38$~eV the sp$\sigma$ hopping is included in the analysis and closes the sp$\sigma$–$\pi$ superexchange path. This induces a strong FM drop (of $\sim-3.5$~meV) in the $d_{z'^{2}},d_{x'y'}$ contribution, seen in Fig.~\ref{fig_CrI3}a.

The second–largest contribution to $J_{1NN}$ of CrI$_{3}$ is the AFM $d_{x'^{2}-y'^{2}},d_{x'^{2}-y'^{2}}$ one. It exhibits a sequence of jumps and drops as different paths are closed with $\tau$ decreased (brown line in Fig.~\ref{fig_CrI3}a), however we will focus on the two most important ones – the FM drop at $\tau = 0.280$~eV, and the AFM jump at $\tau = 0.277$~eV. Those occurred when two different $\pi$ hoppings between $p$ orbitals were included and the super–superexchange paths were closed (shown in panels \textit{iii} and \textit{iv} in Fig.~\ref{fig_closed_paths}a, respectively). The other, less dominant contributions, are detailed in the SM \cite{SM}.

In case of $J_{2NN}$ in CrI$_{3}$, the underlying physics is simpler: all $t_{2g},t_{2g}$ contributions are zero; the most of FM exchange stems from the $d_{x'y'},d_{z'^{2}}$ contribution, that belongs to $t_{2g},e_{g}$ class (and its twin; see green bar and line in Fig.~\ref{fig_CrI3}b), while three $e_{g},e_{g}$ contributions are minor (the other three bars shown in Fig.~\ref{fig_CrI3}b). We thus focus on the origin of the dominant contribution, $d_{z'^{2}},d_{x'y'}$. The first two increments in the green line of Fig.~\ref{fig_CrI3}b, at $\tau = 0.320$~eV and $\tau = 0.287$~eV, are due to the inclusion of spin–up and spin–down $\Lambda$ hopping, which closes the $\sigma$–$\Lambda$–$\pi$ super–superexchange path for spins up and down (see panel \textit{i} in Fig.~\ref{fig_closed_paths}b). The next two increments, at $\tau = 0.189$~eV and $\tau = 0.170$~eV occur due to activation of another spin–up and spin–down $\Lambda$–hopping and the closure of the second $\sigma$–$\Lambda$–$\pi$ super–superexchange path (panel \textit{ii} in Fig.~\ref{fig_closed_paths}b). The other relevant FM increments include processes such as sp$\sigma$–$\pi$–$\pi$ super–superexchange, $\Lambda$–$\pi$ superexchange, and large number of other multi–stage–paths which are closed only when extremely small hopping parameters are included, i.e. for $\tau\approx0.001-0.02$~eV. However, since none of them change the FM behavior of the dominant $d_{z'^{2}},d_{x'y'}$ contribution, we will not discuss them in detail.

\section{The case of NiI$_{2}$} Let us now proceed with the analysis of monolayer NiI$_{2}$, with FM $J_{1NN} = -0.54$~meV and AFM $J_{3NN} = +0.77$~meV. Both 1NN and 3NN pairs of atoms considered are along the global $x$ direction.

\begin{figure}[t]
\centering
\includegraphics[width=\columnwidth]{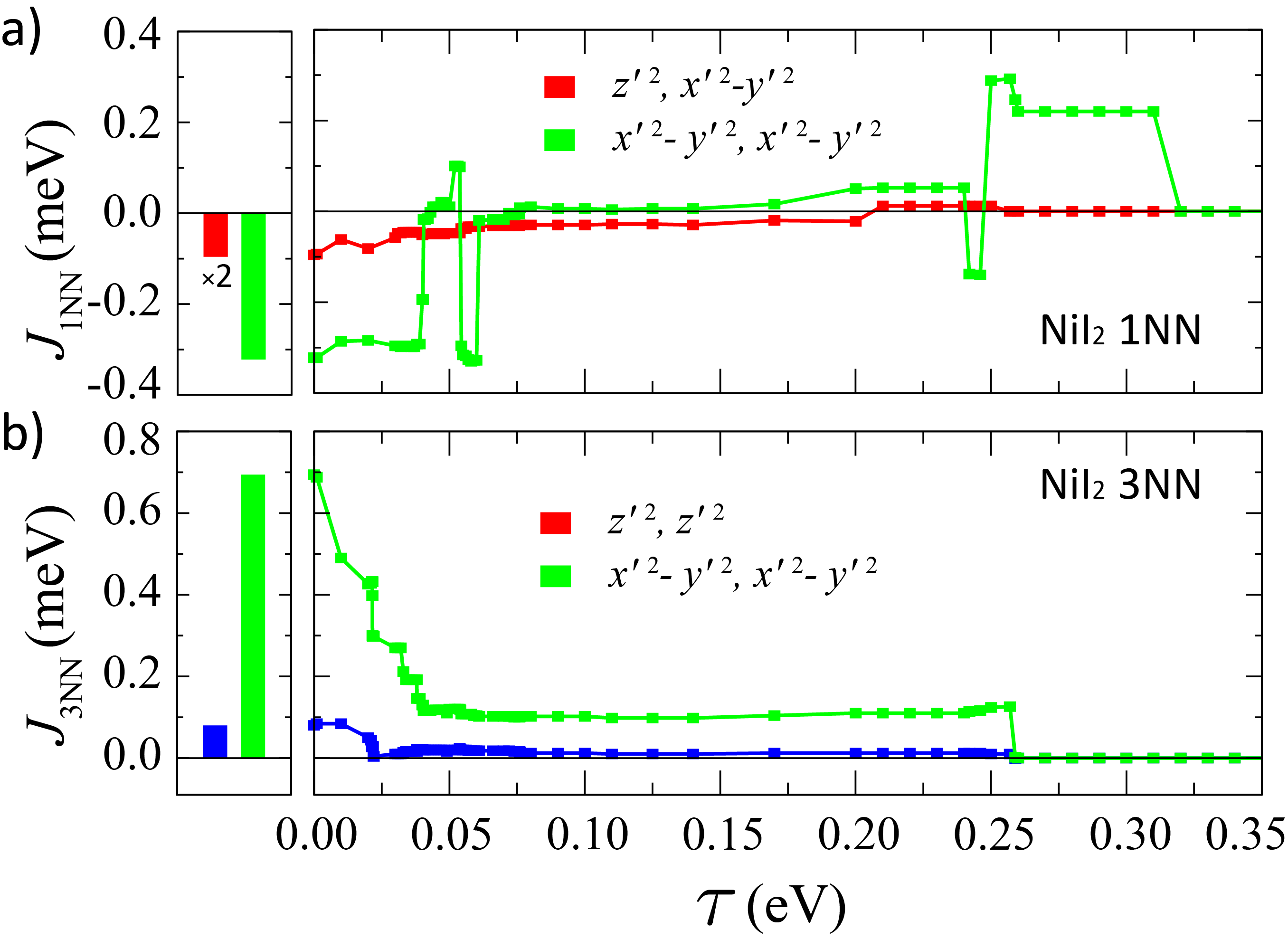}
\caption{\textbf{Origin of $J_{1NN}$ and $J_{3NN}$ in NiI$_{2}$.} (a) Orbitally–resolved contributions to the 1NN magnetic exchange (bars) and their hopping–dependent origin (lines) in monolayer NiI$_{2}$. (b) Same as (a), for the 3NN magnetic exchange.}
\label{fig_NiI2}
\end{figure}

$J_{1NN}$ has three main non–zero orbitally–resolved contributions – all from the $e_{g},e_{g}$ class. The largest part of the net FM 1NN magnetic exchange stems from the (to date underrated) $d_{x'^{2}-y'^{2}},d_{x'^{2}-y'^{2}}$ contribution. In Fig.~\ref{fig_NiI2}a, for $\tau > 0.320$~eV, the $d_{x'^{2}-y'^{2}},d_{x'^{2}-y'^{2}}$ contribution is zero, and the first increment (see the green line) originates from one $\sigma$–$\pi$–$\sigma$ super–superexchange (panel \textit{iii} in Fig.~\ref{fig_closed_paths}a), inducing $+0.22$~meV AFM exchange between 1NN Ni atoms. The FM drop of $-0.24$~meV seen at $\tau = 0.250$~eV is due to another $\sigma$–$\pi$–$\sigma$ super–superexchange (shown in panel \textit{iv} in Fig.~\ref{fig_closed_paths}a) and cancels out the prior AFM contribution. The next FM drop of ($\sim-0.33$~meV) at $\tau = 0.061$~eV is due to direct hopping between the $d_{x'^{2}-y'^{2}}$ orbitals with spin down on the 1NN Ni atoms. The following increment at $\tau = 0.055$~eV is strongly AFM and adds $+0.41$~meV magnetic exchange to the $d_{x'^{2}-y'^{2}},d_{x'^{2}-y'^{2}}$ contribution. Here $F$ and $S$ hoppings are included, which close the “$F$–$S$–$\sigma$” super–superexchange path (panel \textit{v} in Fig.~\ref{fig_closed_paths}a), and cancel out previous FM contribution caused by direct exchange. The final important FM drop was found at $\tau = 0.04$~eV, the value of $F'$ hopping, that closes additional $F'$–$\sigma$ superexchange path (see panel \textit{vi} in Fig.~\ref{fig_closed_paths}a). This sets the $d_{x'^{2}-y'^{2}},d_{x'^{2}-y'^{2}}$ contribution to its final FM value of c.a. $-0.30$~meV. 

In case of AFM $J_{3NN}$ of NiI$_{2}$, there are only two non–zero orbitally–resolved terms contributing, and both of them belong to the $e_{g},e_{g}$ class. As shown in Fig.~\ref{fig_NiI2}b, minor part of $J_{3NN}$ ($+0.08$~meV) originates from $d_{z'^{2}},d_{z'^{2}}$ (blue bar) and major part ($+0.69$~meV) stems from $d_{x'^{2}-y'^{2}},d_{x'^{2}-y'^{2}}$ interaction (green bar). We therefore detail the latter.

The first jump in the green line of Fig.~\ref{fig_NiI2}b is found at $\tau = 0.257$~eV and introduces $+0.12$~meV AFM exchange between 3NN Ni atoms. This is due to inclusion of $\Lambda$ hopping which closes the $\sigma$–$\Lambda$–$\sigma$ super–superexchange path (panel \textit{i} in Fig.~\ref{fig_closed_paths}c). The second pronounced AFM increase of $+0.15$~meV occurs between $\tau = 0.0379$~eV and $\tau = 0.032$~eV and is due to the inclusion of $\pi'$ hopping that closes the $\sigma$–$\pi'$–$\sigma$ super–superexchange path (panel \textit{ii} in Fig.~\ref{fig_closed_paths}c). The third significant AFM jump of $+0.13$~meV is found at $\tau = 0.0216$~eV, due to the inclusion of $LR$ hopping that closes the $LR$–$\sigma$ superexchange path (panel \textit{iii} in Fig.~\ref{fig_closed_paths}c). Latter three processes together yield about $+0.40$~meV AFM 3NN magnetic exchange. The remaining $+0.29$~meV AFM exchange of the $d_{x'^{2}-y'^{2}},d_{x'^{2}-y'^{2}}$ term stems from many additional paths which are closed only when smallest hopping parameters are included ($\tau\approx 0.001 - 0.02$~eV). These paths correspond to higher–order superexchange processes with many intermediate stations between the initial and final Ni atoms. Although weakly contributing individually, these higher–order paths have a sizable cumulative effect. 

\section{Discussion and conclusions}
Having dissected the most important contributions to exchange in two archetypal 2D magnetic materials, we go a step further to provide insight into the importance of electronic occupation and the on–site energies of orbitals for magnetic exchange interactions in the system. 

Let us first artificially impose a $d^{3}$ TM atom in the Hamiltonian corresponding to NiI$_{2}$. In a first approximation, this corresponds to VI$_{2}$, a material that exhibits the triangular TM lattice like NiI$_{2}$ and a $d^{3}$ configuration for each TM atom. In that case, we find the 1NN magnetic interaction to be AFM, which is in agreement with existing predictions for monolayer VI${2}$ in the literature~\cite{vi2_1, vi2_2, vi2_exp}. As in CrI$_{3}$, we again observe the competition between FM $d_{z'^{2}},d_{x'y'}$ and AFM $d_{x'^{2}-y'^{2}},d_{x'^{2}-y'^{2}}$ contributions, and in this case, the AFM contribution prevails. Therefore, for the $d^{3}$ configuration of the TM, there is no universal rule for the sign of $J_{1NN}$, due to the competition between the two comparable terms of opposite sign.

Conversely, we next impose the $d^{8}$ configuration on the TM atom in the Hamiltonian corresponding to CrI$_{3}$. In the simplest approximation, this corresponds to NiPS$_{3}$, a material with a hexagonal TM lattice similar to CrI$_{3}$ and a $d^{8}$ configuration for each TM atom. In this case, the 1NN magnetic interaction is found to be FM, the 2NN one is negligible, and the 3NN magnetic interaction is AFM. This result is indeed consistent with known magnetic exchange data for NiPS$_{3}$~\cite{nips_feps_67, nips_feps_70, nips76, nips79, nips_feps_cops}. Based on these findings, we hypothesize the general rule that the $d^{8}$ configuration always leads to an FM $J_{1NN}$–AFM $J_{3NN}$ exchange combination, regardless of the lattice geometry. The underlying reason for this universal rule in the case of the $d^{8}$ configuration of the TM atom is the absence of competition, namely the $d_{x'^{2}-y'^{2}},d_{x'^{2}-y'^{2}}$ interaction dominates the physics of both FM $J_{1NN}$ and AFM $J_{3NN}$.

Furthermore, based on available literature, this rule appears to be even more general. Namely, the same FM $J_{1NN}$–AFM $J_{3NN}$ physics is frequently reported in FePS$_{3}$~\cite{nips_feps_67, nips_feps_70, feps77, feps78, nips_feps_cops} and CoPS$_{3}$~\cite{cops_68, cops_75, nips_feps_cops}, which have $d^{6}$ and $d^{7}$ electronic configurations of the TM atoms. We hypothesize that the reason is the full occupation of one or more $t_{2g}$ orbitals in the case of more than 5 electrons in $d$–shell, making $t_{2g}$ orbitals inactive and leaving the control of magnetic exchange to $e_{g},e_{g}$ terms. 

To summarize, we have established a general and objective microscopic framework for dissecting the physics of magnetic interactions in a given material of any dimensionality. Through the examples of CrI$_{3}$ and NiI$_{2}$ monolayers, the archetypal magnetic 2D materials, we highlighted the critical role of higher–order superexchange interactions in magnetic exchange between transition–metal atoms, even when the involved $d$ orbitals are nominally unoccupied. Furthermore, we demonstrated how the electronic configurations of transition–metal atoms shape the magnetic exchange interactions, and outlined cases where universal rules can be established. These findings clearly exemplify the importance of understanding magnetism at the orbitally–resolved level, and call for further such analysis in all systems of interest. 

\begin{acknowledgments}
This work was supported by the Research Foundation–Flanders (FWO–Vl) and the FWO–FNRS EOS project ShapeME. D.\v{S}. is a doctoral fellow of FWO–Vl under contract No. 11J4322N. C.B. is a senior post–doctoral fellow of FWO–Vl under contract No. 12E8823N. The computational resources and services for this work were provided by the VSC (Flemish Supercomputer Center), funded by the FWO and the Flemish Government – department EWI.
\end{acknowledgments}

\bibliographystyle{apsrev4-2}
\bibliography{ref}

\begin{thebibliography}{75}%
\makeatletter
\providecommand \@ifxundefined [1]{%
 \@ifx{#1\undefined}
}%
\providecommand \@ifnum [1]{%
 \ifnum #1\expandafter \@firstoftwo
 \else \expandafter \@secondoftwo
 \fi
}%
\providecommand \@ifx [1]{%
 \ifx #1\expandafter \@firstoftwo
 \else \expandafter \@secondoftwo
 \fi
}%
\providecommand \natexlab [1]{#1}%
\providecommand \enquote  [1]{``#1''}%
\providecommand \bibnamefont  [1]{#1}%
\providecommand \bibfnamefont [1]{#1}%
\providecommand \citenamefont [1]{#1}%
\providecommand \href@noop [0]{\@secondoftwo}%
\providecommand \href [0]{\begingroup \@sanitize@url \@href}%
\providecommand \@href[1]{\@@startlink{#1}\@@href}%
\providecommand \@@href[1]{\endgroup#1\@@endlink}%
\providecommand \@sanitize@url [0]{\catcode `\\12\catcode `\$12\catcode `\&12\catcode `\#12\catcode `\^12\catcode `\_12\catcode `\%12\relax}%
\providecommand \@@startlink[1]{}%
\providecommand \@@endlink[0]{}%
\providecommand \url  [0]{\begingroup\@sanitize@url \@url }%
\providecommand \@url [1]{\endgroup\@href {#1}{\urlprefix }}%
\providecommand \urlprefix  [0]{URL }%
\providecommand \Eprint [0]{\href }%
\providecommand \doibase [0]{https://doi.org/}%
\providecommand \selectlanguage [0]{\@gobble}%
\providecommand \bibinfo  [0]{\@secondoftwo}%
\providecommand \bibfield  [0]{\@secondoftwo}%
\providecommand \translation [1]{[#1]}%
\providecommand \BibitemOpen [0]{}%
\providecommand \bibitemStop [0]{}%
\providecommand \bibitemNoStop [0]{.\EOS\space}%
\providecommand \EOS [0]{\spacefactor3000\relax}%
\providecommand \BibitemShut  [1]{\csname bibitem#1\endcsname}%
\let\auto@bib@innerbib\@empty
\bibitem [{\citenamefont {Huang}\ \emph {et~al.}(2017)\citenamefont {Huang}, \citenamefont {Clark}, \citenamefont {Navarro-Moratalla}, \citenamefont {Klein}, \citenamefont {Cheng}, \citenamefont {Seyler}, \citenamefont {Zhong}, \citenamefont {Schmidgall}, \citenamefont {McGuire}, \citenamefont {Cobden}, \citenamefont {Yao}, \citenamefont {Xiao}, \citenamefont {Jarillo-Herrero},\ and\ \citenamefont {Xu}}]{CrI32017}%
  \BibitemOpen
  \bibfield  {author} {\bibinfo {author} {\bibfnamefont {B.}~\bibnamefont {Huang}}, \bibinfo {author} {\bibfnamefont {G.}~\bibnamefont {Clark}}, \bibinfo {author} {\bibfnamefont {E.}~\bibnamefont {Navarro-Moratalla}}, \bibinfo {author} {\bibfnamefont {D.~R.}\ \bibnamefont {Klein}}, \bibinfo {author} {\bibfnamefont {R.}~\bibnamefont {Cheng}}, \bibinfo {author} {\bibfnamefont {K.~L.}\ \bibnamefont {Seyler}}, \bibinfo {author} {\bibfnamefont {D.}~\bibnamefont {Zhong}}, \bibinfo {author} {\bibfnamefont {E.}~\bibnamefont {Schmidgall}}, \bibinfo {author} {\bibfnamefont {M.~A.}\ \bibnamefont {McGuire}}, \bibinfo {author} {\bibfnamefont {D.~H.}\ \bibnamefont {Cobden}}, \bibinfo {author} {\bibfnamefont {W.}~\bibnamefont {Yao}}, \bibinfo {author} {\bibfnamefont {D.}~\bibnamefont {Xiao}}, \bibinfo {author} {\bibfnamefont {P.}~\bibnamefont {Jarillo-Herrero}},\ and\ \bibinfo {author} {\bibfnamefont {X.}~\bibnamefont {Xu}},\ }\href {https://doi.org/10.1038/nature22391} {\bibfield  {journal} {\bibinfo  {journal} {Nature}\
  }\textbf {\bibinfo {volume} {546}},\ \bibinfo {pages} {270} (\bibinfo {year} {2017})}\BibitemShut {NoStop}%
\bibitem [{\citenamefont {Gong}\ \emph {et~al.}(2017)\citenamefont {Gong}, \citenamefont {Li}, \citenamefont {Li}, \citenamefont {Ji}, \citenamefont {Stern}, \citenamefont {Xia}, \citenamefont {Cao}, \citenamefont {Bao}, \citenamefont {Wang}, \citenamefont {Wang}, \citenamefont {Qiu}, \citenamefont {Cava}, \citenamefont {Louie}, \citenamefont {Xia},\ and\ \citenamefont {Zhang}}]{CrGeTe2017}%
  \BibitemOpen
  \bibfield  {author} {\bibinfo {author} {\bibfnamefont {C.}~\bibnamefont {Gong}}, \bibinfo {author} {\bibfnamefont {L.}~\bibnamefont {Li}}, \bibinfo {author} {\bibfnamefont {Z.}~\bibnamefont {Li}}, \bibinfo {author} {\bibfnamefont {H.}~\bibnamefont {Ji}}, \bibinfo {author} {\bibfnamefont {A.}~\bibnamefont {Stern}}, \bibinfo {author} {\bibfnamefont {Y.}~\bibnamefont {Xia}}, \bibinfo {author} {\bibfnamefont {T.}~\bibnamefont {Cao}}, \bibinfo {author} {\bibfnamefont {W.}~\bibnamefont {Bao}}, \bibinfo {author} {\bibfnamefont {C.}~\bibnamefont {Wang}}, \bibinfo {author} {\bibfnamefont {Y.}~\bibnamefont {Wang}}, \bibinfo {author} {\bibfnamefont {Z.~Q.}\ \bibnamefont {Qiu}}, \bibinfo {author} {\bibfnamefont {R.~J.}\ \bibnamefont {Cava}}, \bibinfo {author} {\bibfnamefont {S.~G.}\ \bibnamefont {Louie}}, \bibinfo {author} {\bibfnamefont {J.}~\bibnamefont {Xia}},\ and\ \bibinfo {author} {\bibfnamefont {X.}~\bibnamefont {Zhang}},\ }\href {https://doi.org/10.1038/nature22060} {\bibfield  {journal} {\bibinfo  {journal}
  {Nature}\ }\textbf {\bibinfo {volume} {546}},\ \bibinfo {pages} {265} (\bibinfo {year} {2017})}\BibitemShut {NoStop}%
\bibitem [{\citenamefont {Mermin}\ and\ \citenamefont {Wagner}(1966)}]{MerminWagner}%
  \BibitemOpen
  \bibfield  {author} {\bibinfo {author} {\bibfnamefont {N.~D.}\ \bibnamefont {Mermin}}\ and\ \bibinfo {author} {\bibfnamefont {H.}~\bibnamefont {Wagner}},\ }\href {https://doi.org/10.1103/PhysRevLett.17.1133} {\bibfield  {journal} {\bibinfo  {journal} {Phys. Rev. Lett.}\ }\textbf {\bibinfo {volume} {17}},\ \bibinfo {pages} {1133} (\bibinfo {year} {1966})}\BibitemShut {NoStop}%
\bibitem [{\citenamefont {Xing}\ \emph {et~al.}(2017)\citenamefont {Xing}, \citenamefont {Chen}, \citenamefont {Odenthal}, \citenamefont {Zhang}, \citenamefont {Yuan}, \citenamefont {Su}, \citenamefont {Song}, \citenamefont {Wang}, \citenamefont {Zhong}, \citenamefont {Jia}, \citenamefont {Xie}, \citenamefont {Li},\ and\ \citenamefont {Han}}]{CrGeTe2017_Xing}%
  \BibitemOpen
  \bibfield  {author} {\bibinfo {author} {\bibfnamefont {W.}~\bibnamefont {Xing}}, \bibinfo {author} {\bibfnamefont {Y.}~\bibnamefont {Chen}}, \bibinfo {author} {\bibfnamefont {P.~M.}\ \bibnamefont {Odenthal}}, \bibinfo {author} {\bibfnamefont {X.}~\bibnamefont {Zhang}}, \bibinfo {author} {\bibfnamefont {W.}~\bibnamefont {Yuan}}, \bibinfo {author} {\bibfnamefont {T.}~\bibnamefont {Su}}, \bibinfo {author} {\bibfnamefont {Q.}~\bibnamefont {Song}}, \bibinfo {author} {\bibfnamefont {T.}~\bibnamefont {Wang}}, \bibinfo {author} {\bibfnamefont {J.}~\bibnamefont {Zhong}}, \bibinfo {author} {\bibfnamefont {S.}~\bibnamefont {Jia}}, \bibinfo {author} {\bibfnamefont {X.~C.}\ \bibnamefont {Xie}}, \bibinfo {author} {\bibfnamefont {Y.}~\bibnamefont {Li}},\ and\ \bibinfo {author} {\bibfnamefont {W.}~\bibnamefont {Han}},\ }\href {https://doi.org/10.1088/2053-1583/aa7034} {\bibfield  {journal} {\bibinfo  {journal} {2D Materials}\ }\textbf {\bibinfo {volume} {4}},\ \bibinfo {pages} {024009} (\bibinfo {year} {2017})}\BibitemShut
  {NoStop}%
\bibitem [{\citenamefont {Chen}\ \emph {et~al.}(2017)\citenamefont {Chen}, \citenamefont {Xing}, \citenamefont {Wang}, \citenamefont {Shen}, \citenamefont {Yuan}, \citenamefont {Su}, \citenamefont {Ma}, \citenamefont {Yao}, \citenamefont {Zhong}, \citenamefont {Yun}, \citenamefont {Xie}, \citenamefont {Jia},\ and\ \citenamefont {Han}}]{CrGeTe2017_Chen}%
  \BibitemOpen
  \bibfield  {author} {\bibinfo {author} {\bibfnamefont {Y.}~\bibnamefont {Chen}}, \bibinfo {author} {\bibfnamefont {W.}~\bibnamefont {Xing}}, \bibinfo {author} {\bibfnamefont {X.}~\bibnamefont {Wang}}, \bibinfo {author} {\bibfnamefont {B.}~\bibnamefont {Shen}}, \bibinfo {author} {\bibfnamefont {W.}~\bibnamefont {Yuan}}, \bibinfo {author} {\bibfnamefont {T.}~\bibnamefont {Su}}, \bibinfo {author} {\bibfnamefont {Y.}~\bibnamefont {Ma}}, \bibinfo {author} {\bibfnamefont {Y.}~\bibnamefont {Yao}}, \bibinfo {author} {\bibfnamefont {J.}~\bibnamefont {Zhong}}, \bibinfo {author} {\bibfnamefont {Y.}~\bibnamefont {Yun}}, \bibinfo {author} {\bibfnamefont {X.~C.}\ \bibnamefont {Xie}}, \bibinfo {author} {\bibfnamefont {S.}~\bibnamefont {Jia}},\ and\ \bibinfo {author} {\bibfnamefont {W.}~\bibnamefont {Han}},\ }\href {https://doi.org/10.1021/acsami.7b14795} {\bibfield  {journal} {\bibinfo  {journal} {ACS Appl. Mater. Interfaces}\ }\textbf {\bibinfo {volume} {10}},\ \bibinfo {pages} {1383} (\bibinfo {year}
  {2017})}\BibitemShut {NoStop}%
\bibitem [{\citenamefont {Bonilla}\ \emph {et~al.}(2018)\citenamefont {Bonilla}, \citenamefont {Kolekar}, \citenamefont {Ma}, \citenamefont {Diaz}, \citenamefont {Kalappattil}, \citenamefont {Das}, \citenamefont {Eggers}, \citenamefont {Gutierrez}, \citenamefont {Phan},\ and\ \citenamefont {Batzill}}]{VSe2}%
  \BibitemOpen
  \bibfield  {author} {\bibinfo {author} {\bibfnamefont {M.}~\bibnamefont {Bonilla}}, \bibinfo {author} {\bibfnamefont {S.}~\bibnamefont {Kolekar}}, \bibinfo {author} {\bibfnamefont {Y.}~\bibnamefont {Ma}}, \bibinfo {author} {\bibfnamefont {H.~C.}\ \bibnamefont {Diaz}}, \bibinfo {author} {\bibfnamefont {V.}~\bibnamefont {Kalappattil}}, \bibinfo {author} {\bibfnamefont {R.}~\bibnamefont {Das}}, \bibinfo {author} {\bibfnamefont {T.}~\bibnamefont {Eggers}}, \bibinfo {author} {\bibfnamefont {H.~R.}\ \bibnamefont {Gutierrez}}, \bibinfo {author} {\bibfnamefont {M.-H.}\ \bibnamefont {Phan}},\ and\ \bibinfo {author} {\bibfnamefont {M.}~\bibnamefont {Batzill}},\ }\href {https://doi.org/10.1038/s41565-018-0063-9} {\bibfield  {journal} {\bibinfo  {journal} {Nature Nanotechnology}\ }\textbf {\bibinfo {volume} {13}},\ \bibinfo {pages} {289} (\bibinfo {year} {2018})}\BibitemShut {NoStop}%
\bibitem [{\citenamefont {Jiang}\ \emph {et~al.}(2018)\citenamefont {Jiang}, \citenamefont {Shan},\ and\ \citenamefont {Mak}}]{Efield_switch}%
  \BibitemOpen
  \bibfield  {author} {\bibinfo {author} {\bibfnamefont {S.}~\bibnamefont {Jiang}}, \bibinfo {author} {\bibfnamefont {J.}~\bibnamefont {Shan}},\ and\ \bibinfo {author} {\bibfnamefont {K.~F.}\ \bibnamefont {Mak}},\ }\href {https://doi.org/10.1038/s41563-018-0040-6} {\bibfield  {journal} {\bibinfo  {journal} {Nature Materials}\ }\textbf {\bibinfo {volume} {17}},\ \bibinfo {pages} {406} (\bibinfo {year} {2018})}\BibitemShut {NoStop}%
\bibitem [{\citenamefont {O’Hara}\ \emph {et~al.}(2018{\natexlab{a}})\citenamefont {O’Hara}, \citenamefont {Zhu}, \citenamefont {Trout}, \citenamefont {Ahmed}, \citenamefont {Luo}, \citenamefont {Lee}, \citenamefont {Brenner}, \citenamefont {Rajan}, \citenamefont {Gupta}, \citenamefont {McComb},\ and\ \citenamefont {Kawakami}}]{MnSex}%
  \BibitemOpen
  \bibfield  {author} {\bibinfo {author} {\bibfnamefont {D.~J.}\ \bibnamefont {O’Hara}}, \bibinfo {author} {\bibfnamefont {T.}~\bibnamefont {Zhu}}, \bibinfo {author} {\bibfnamefont {A.~H.}\ \bibnamefont {Trout}}, \bibinfo {author} {\bibfnamefont {A.~S.}\ \bibnamefont {Ahmed}}, \bibinfo {author} {\bibfnamefont {Y.~K.}\ \bibnamefont {Luo}}, \bibinfo {author} {\bibfnamefont {C.~H.}\ \bibnamefont {Lee}}, \bibinfo {author} {\bibfnamefont {M.~R.}\ \bibnamefont {Brenner}}, \bibinfo {author} {\bibfnamefont {S.}~\bibnamefont {Rajan}}, \bibinfo {author} {\bibfnamefont {J.~A.}\ \bibnamefont {Gupta}}, \bibinfo {author} {\bibfnamefont {D.~W.}\ \bibnamefont {McComb}},\ and\ \bibinfo {author} {\bibfnamefont {R.~K.}\ \bibnamefont {Kawakami}},\ }\href {https://doi.org/10.1021/acs.nanolett.8b00683} {\bibfield  {journal} {\bibinfo  {journal} {Nano Letters}\ }\textbf {\bibinfo {volume} {18}},\ \bibinfo {pages} {3125} (\bibinfo {year} {2018}{\natexlab{a}})},\ \bibinfo {note} {pMID: 29608316}\BibitemShut {NoStop}%
\bibitem [{\citenamefont {Abramchuk}\ \emph {et~al.}(2018)\citenamefont {Abramchuk}, \citenamefont {Jaszewski}, \citenamefont {Metz}, \citenamefont {Osterhoudt}, \citenamefont {Wang}, \citenamefont {Burch},\ and\ \citenamefont {Tafti}}]{CrCl3-xBr3}%
  \BibitemOpen
  \bibfield  {author} {\bibinfo {author} {\bibfnamefont {M.}~\bibnamefont {Abramchuk}}, \bibinfo {author} {\bibfnamefont {S.}~\bibnamefont {Jaszewski}}, \bibinfo {author} {\bibfnamefont {K.~R.}\ \bibnamefont {Metz}}, \bibinfo {author} {\bibfnamefont {G.~B.}\ \bibnamefont {Osterhoudt}}, \bibinfo {author} {\bibfnamefont {Y.}~\bibnamefont {Wang}}, \bibinfo {author} {\bibfnamefont {K.~S.}\ \bibnamefont {Burch}},\ and\ \bibinfo {author} {\bibfnamefont {F.}~\bibnamefont {Tafti}},\ }\href {https://doi.org/10.1002/adma.201801325} {\bibfield  {journal} {\bibinfo  {journal} {Advanced Materials}\ }\textbf {\bibinfo {volume} {30}},\ \bibinfo {pages} {1801325} (\bibinfo {year} {2018})}\BibitemShut {NoStop}%
\bibitem [{\citenamefont {Song}\ \emph {et~al.}(2018)\citenamefont {Song}, \citenamefont {Cai}, \citenamefont {Tu}, \citenamefont {Zhang}, \citenamefont {Huang}, \citenamefont {Wilson}, \citenamefont {Seyler}, \citenamefont {Zhu}, \citenamefont {Taniguchi}, \citenamefont {Watanabe}, \citenamefont {McGuire}, \citenamefont {Cobden}, \citenamefont {Xiao}, \citenamefont {Yao},\ and\ \citenamefont {Xu}}]{Giant_tunnel}%
  \BibitemOpen
  \bibfield  {author} {\bibinfo {author} {\bibfnamefont {T.}~\bibnamefont {Song}}, \bibinfo {author} {\bibfnamefont {X.}~\bibnamefont {Cai}}, \bibinfo {author} {\bibfnamefont {M.~W.-Y.}\ \bibnamefont {Tu}}, \bibinfo {author} {\bibfnamefont {X.}~\bibnamefont {Zhang}}, \bibinfo {author} {\bibfnamefont {B.}~\bibnamefont {Huang}}, \bibinfo {author} {\bibfnamefont {N.~P.}\ \bibnamefont {Wilson}}, \bibinfo {author} {\bibfnamefont {K.~L.}\ \bibnamefont {Seyler}}, \bibinfo {author} {\bibfnamefont {L.}~\bibnamefont {Zhu}}, \bibinfo {author} {\bibfnamefont {T.}~\bibnamefont {Taniguchi}}, \bibinfo {author} {\bibfnamefont {K.}~\bibnamefont {Watanabe}}, \bibinfo {author} {\bibfnamefont {M.~A.}\ \bibnamefont {McGuire}}, \bibinfo {author} {\bibfnamefont {D.~H.}\ \bibnamefont {Cobden}}, \bibinfo {author} {\bibfnamefont {D.}~\bibnamefont {Xiao}}, \bibinfo {author} {\bibfnamefont {W.}~\bibnamefont {Yao}},\ and\ \bibinfo {author} {\bibfnamefont {X.}~\bibnamefont {Xu}},\ }\href {https://doi.org/10.1126/science.aar4851}
  {\bibfield  {journal} {\bibinfo  {journal} {Science}\ }\textbf {\bibinfo {volume} {360}},\ \bibinfo {pages} {1214} (\bibinfo {year} {2018})}\BibitemShut {NoStop}%
\bibitem [{\citenamefont {Klein}\ \emph {et~al.}(2018)\citenamefont {Klein}, \citenamefont {MacNeill}, \citenamefont {Lado}, \citenamefont {Soriano}, \citenamefont {Navarro-Moratalla}, \citenamefont {Watanabe}, \citenamefont {Taniguchi}, \citenamefont {Manni}, \citenamefont {Canfield}, \citenamefont {Fernández-Rossier},\ and\ \citenamefont {Jarillo-Herrero}}]{2DvdW_etunnel}%
  \BibitemOpen
  \bibfield  {author} {\bibinfo {author} {\bibfnamefont {D.~R.}\ \bibnamefont {Klein}}, \bibinfo {author} {\bibfnamefont {D.}~\bibnamefont {MacNeill}}, \bibinfo {author} {\bibfnamefont {J.~L.}\ \bibnamefont {Lado}}, \bibinfo {author} {\bibfnamefont {D.}~\bibnamefont {Soriano}}, \bibinfo {author} {\bibfnamefont {E.}~\bibnamefont {Navarro-Moratalla}}, \bibinfo {author} {\bibfnamefont {K.}~\bibnamefont {Watanabe}}, \bibinfo {author} {\bibfnamefont {T.}~\bibnamefont {Taniguchi}}, \bibinfo {author} {\bibfnamefont {S.}~\bibnamefont {Manni}}, \bibinfo {author} {\bibfnamefont {P.}~\bibnamefont {Canfield}}, \bibinfo {author} {\bibfnamefont {J.}~\bibnamefont {Fernández-Rossier}},\ and\ \bibinfo {author} {\bibfnamefont {P.}~\bibnamefont {Jarillo-Herrero}},\ }\href {https://doi.org/10.1126/science.aar3617} {\bibfield  {journal} {\bibinfo  {journal} {Science}\ }\textbf {\bibinfo {volume} {360}},\ \bibinfo {pages} {1218} (\bibinfo {year} {2018})}\BibitemShut {NoStop}%
\bibitem [{\citenamefont {Kim}\ \emph {et~al.}(2018)\citenamefont {Kim}, \citenamefont {Yang}, \citenamefont {Patel}, \citenamefont {Sfigakis}, \citenamefont {Li}, \citenamefont {Tian}, \citenamefont {Lei},\ and\ \citenamefont {Tsen}}]{onemilion}%
  \BibitemOpen
  \bibfield  {author} {\bibinfo {author} {\bibfnamefont {H.~H.}\ \bibnamefont {Kim}}, \bibinfo {author} {\bibfnamefont {B.}~\bibnamefont {Yang}}, \bibinfo {author} {\bibfnamefont {T.}~\bibnamefont {Patel}}, \bibinfo {author} {\bibfnamefont {F.}~\bibnamefont {Sfigakis}}, \bibinfo {author} {\bibfnamefont {C.}~\bibnamefont {Li}}, \bibinfo {author} {\bibfnamefont {S.}~\bibnamefont {Tian}}, \bibinfo {author} {\bibfnamefont {H.}~\bibnamefont {Lei}},\ and\ \bibinfo {author} {\bibfnamefont {A.~W.}\ \bibnamefont {Tsen}},\ }\href {https://doi.org/10.1021/acs.nanolett.8b01552} {\bibfield  {journal} {\bibinfo  {journal} {Nano Letters}\ }\textbf {\bibinfo {volume} {18}},\ \bibinfo {pages} {4885} (\bibinfo {year} {2018})}\BibitemShut {NoStop}%
\bibitem [{\citenamefont {Deng}\ \emph {et~al.}(2018)\citenamefont {Deng}, \citenamefont {Yu}, \citenamefont {Song}, \citenamefont {Zhang}, \citenamefont {Wang}, \citenamefont {Sun}, \citenamefont {Yi}, \citenamefont {Wu}, \citenamefont {Wu}, \citenamefont {Zhu}, \citenamefont {Wang}, \citenamefont {Chen},\ and\ \citenamefont {Zhang}}]{Deng2018_FeGeTe}%
  \BibitemOpen
  \bibfield  {author} {\bibinfo {author} {\bibfnamefont {Y.}~\bibnamefont {Deng}}, \bibinfo {author} {\bibfnamefont {Y.}~\bibnamefont {Yu}}, \bibinfo {author} {\bibfnamefont {Y.}~\bibnamefont {Song}}, \bibinfo {author} {\bibfnamefont {J.}~\bibnamefont {Zhang}}, \bibinfo {author} {\bibfnamefont {N.~Z.}\ \bibnamefont {Wang}}, \bibinfo {author} {\bibfnamefont {Z.}~\bibnamefont {Sun}}, \bibinfo {author} {\bibfnamefont {Y.}~\bibnamefont {Yi}}, \bibinfo {author} {\bibfnamefont {Y.~Z.}\ \bibnamefont {Wu}}, \bibinfo {author} {\bibfnamefont {S.}~\bibnamefont {Wu}}, \bibinfo {author} {\bibfnamefont {J.}~\bibnamefont {Zhu}}, \bibinfo {author} {\bibfnamefont {J.}~\bibnamefont {Wang}}, \bibinfo {author} {\bibfnamefont {X.~H.}\ \bibnamefont {Chen}},\ and\ \bibinfo {author} {\bibfnamefont {Y.}~\bibnamefont {Zhang}},\ }\href {https://doi.org/10.1038/s41586-018-0626-9} {\bibfield  {journal} {\bibinfo  {journal} {Nature}\ }\textbf {\bibinfo {volume} {563}},\ \bibinfo {pages} {94} (\bibinfo {year} {2018})}\BibitemShut {NoStop}%
\bibitem [{\citenamefont {Sivadas}\ \emph {et~al.}(2018)\citenamefont {Sivadas}, \citenamefont {Okamoto}, \citenamefont {Xu}, \citenamefont {Fennie},\ and\ \citenamefont {Xiao}}]{sivadas2018}%
  \BibitemOpen
  \bibfield  {author} {\bibinfo {author} {\bibfnamefont {N.}~\bibnamefont {Sivadas}}, \bibinfo {author} {\bibfnamefont {S.}~\bibnamefont {Okamoto}}, \bibinfo {author} {\bibfnamefont {X.}~\bibnamefont {Xu}}, \bibinfo {author} {\bibfnamefont {C.~J.}\ \bibnamefont {Fennie}},\ and\ \bibinfo {author} {\bibfnamefont {D.}~\bibnamefont {Xiao}},\ }\href {https://doi.org/10.1021/acs.nanolett.8b03321} {\bibfield  {journal} {\bibinfo  {journal} {Nano Letters}\ }\textbf {\bibinfo {volume} {18}},\ \bibinfo {pages} {7658} (\bibinfo {year} {2018})}\BibitemShut {NoStop}%
\bibitem [{\citenamefont {Song}\ \emph {et~al.}(2019)\citenamefont {Song}, \citenamefont {Fei}, \citenamefont {Yankowitz}, \citenamefont {Lin}, \citenamefont {Jiang}, \citenamefont {Hwangbo}, \citenamefont {Zhang}, \citenamefont {Sun}, \citenamefont {Taniguchi}, \citenamefont {Watanabe}, \citenamefont {McGuire}, \citenamefont {Graf}, \citenamefont {Cao}, \citenamefont {Chu}, \citenamefont {Cobden}, \citenamefont {Dean}, \citenamefont {Xiao},\ and\ \citenamefont {Xu}}]{song2019}%
  \BibitemOpen
  \bibfield  {author} {\bibinfo {author} {\bibfnamefont {T.}~\bibnamefont {Song}}, \bibinfo {author} {\bibfnamefont {Z.}~\bibnamefont {Fei}}, \bibinfo {author} {\bibfnamefont {M.}~\bibnamefont {Yankowitz}}, \bibinfo {author} {\bibfnamefont {Z.}~\bibnamefont {Lin}}, \bibinfo {author} {\bibfnamefont {Q.}~\bibnamefont {Jiang}}, \bibinfo {author} {\bibfnamefont {K.}~\bibnamefont {Hwangbo}}, \bibinfo {author} {\bibfnamefont {Q.}~\bibnamefont {Zhang}}, \bibinfo {author} {\bibfnamefont {B.}~\bibnamefont {Sun}}, \bibinfo {author} {\bibfnamefont {T.}~\bibnamefont {Taniguchi}}, \bibinfo {author} {\bibfnamefont {K.}~\bibnamefont {Watanabe}}, \bibinfo {author} {\bibfnamefont {M.~A.}\ \bibnamefont {McGuire}}, \bibinfo {author} {\bibfnamefont {D.}~\bibnamefont {Graf}}, \bibinfo {author} {\bibfnamefont {T.}~\bibnamefont {Cao}}, \bibinfo {author} {\bibfnamefont {J.-H.}\ \bibnamefont {Chu}}, \bibinfo {author} {\bibfnamefont {D.~H.}\ \bibnamefont {Cobden}}, \bibinfo {author} {\bibfnamefont {C.~R.}\ \bibnamefont {Dean}},
  \bibinfo {author} {\bibfnamefont {D.}~\bibnamefont {Xiao}},\ and\ \bibinfo {author} {\bibfnamefont {X.}~\bibnamefont {Xu}},\ }\href {https://doi.org/10.1038/s41563-019-0505-2} {\bibfield  {journal} {\bibinfo  {journal} {Nature Materials}\ }\textbf {\bibinfo {volume} {18}},\ \bibinfo {pages} {1298} (\bibinfo {year} {2019})}\BibitemShut {NoStop}%
\bibitem [{\citenamefont {Thiel}\ \emph {et~al.}(2019)\citenamefont {Thiel}, \citenamefont {Wang}, \citenamefont {Tschudin}, \citenamefont {Rohner}, \citenamefont {Gutiérrez-Lezama}, \citenamefont {Ubrig}, \citenamefont {Gibertini}, \citenamefont {Giannini}, \citenamefont {Morpurgo},\ and\ \citenamefont {Maletinsky}}]{thiel2019}%
  \BibitemOpen
  \bibfield  {author} {\bibinfo {author} {\bibfnamefont {L.}~\bibnamefont {Thiel}}, \bibinfo {author} {\bibfnamefont {Z.}~\bibnamefont {Wang}}, \bibinfo {author} {\bibfnamefont {M.~A.}\ \bibnamefont {Tschudin}}, \bibinfo {author} {\bibfnamefont {D.}~\bibnamefont {Rohner}}, \bibinfo {author} {\bibfnamefont {I.}~\bibnamefont {Gutiérrez-Lezama}}, \bibinfo {author} {\bibfnamefont {N.}~\bibnamefont {Ubrig}}, \bibinfo {author} {\bibfnamefont {M.}~\bibnamefont {Gibertini}}, \bibinfo {author} {\bibfnamefont {E.}~\bibnamefont {Giannini}}, \bibinfo {author} {\bibfnamefont {A.~F.}\ \bibnamefont {Morpurgo}},\ and\ \bibinfo {author} {\bibfnamefont {P.}~\bibnamefont {Maletinsky}},\ }\href {https://doi.org/10.1126/science.aav6926} {\bibfield  {journal} {\bibinfo  {journal} {Science}\ }\textbf {\bibinfo {volume} {364}},\ \bibinfo {pages} {973} (\bibinfo {year} {2019})}\BibitemShut {NoStop}%
\bibitem [{\citenamefont {Li}\ \emph {et~al.}(2019)\citenamefont {Li}, \citenamefont {Jiang}, \citenamefont {Sivadas}, \citenamefont {Wang}, \citenamefont {Xu}, \citenamefont {Weber}, \citenamefont {Goldberger}, \citenamefont {Watanabe}, \citenamefont {Taniguchi}, \citenamefont {Fennie}, \citenamefont {Fai~Mak},\ and\ \citenamefont {Shan}}]{li2019}%
  \BibitemOpen
  \bibfield  {author} {\bibinfo {author} {\bibfnamefont {T.}~\bibnamefont {Li}}, \bibinfo {author} {\bibfnamefont {S.}~\bibnamefont {Jiang}}, \bibinfo {author} {\bibfnamefont {N.}~\bibnamefont {Sivadas}}, \bibinfo {author} {\bibfnamefont {Z.}~\bibnamefont {Wang}}, \bibinfo {author} {\bibfnamefont {Y.}~\bibnamefont {Xu}}, \bibinfo {author} {\bibfnamefont {D.}~\bibnamefont {Weber}}, \bibinfo {author} {\bibfnamefont {J.~E.}\ \bibnamefont {Goldberger}}, \bibinfo {author} {\bibfnamefont {K.}~\bibnamefont {Watanabe}}, \bibinfo {author} {\bibfnamefont {T.}~\bibnamefont {Taniguchi}}, \bibinfo {author} {\bibfnamefont {C.~J.}\ \bibnamefont {Fennie}}, \bibinfo {author} {\bibfnamefont {K.}~\bibnamefont {Fai~Mak}},\ and\ \bibinfo {author} {\bibfnamefont {J.}~\bibnamefont {Shan}},\ }\href {https://doi.org/10.1038/s41563-019-0506-1} {\bibfield  {journal} {\bibinfo  {journal} {Nature Materials}\ }\textbf {\bibinfo {volume} {18}},\ \bibinfo {pages} {1303} (\bibinfo {year} {2019})}\BibitemShut {NoStop}%
\bibitem [{\citenamefont {Song}\ \emph {et~al.}(2021)\citenamefont {Song}, \citenamefont {Sun}, \citenamefont {Anderson}, \citenamefont {Wang}, \citenamefont {Qian}, \citenamefont {Taniguchi}, \citenamefont {Watanabe}, \citenamefont {McGuire}, \citenamefont {Stöhr}, \citenamefont {Xiao}, \citenamefont {Cao}, \citenamefont {Wrachtrup},\ and\ \citenamefont {Xu}}]{song2021}%
  \BibitemOpen
  \bibfield  {author} {\bibinfo {author} {\bibfnamefont {T.}~\bibnamefont {Song}}, \bibinfo {author} {\bibfnamefont {Q.-C.}\ \bibnamefont {Sun}}, \bibinfo {author} {\bibfnamefont {E.}~\bibnamefont {Anderson}}, \bibinfo {author} {\bibfnamefont {C.}~\bibnamefont {Wang}}, \bibinfo {author} {\bibfnamefont {J.}~\bibnamefont {Qian}}, \bibinfo {author} {\bibfnamefont {T.}~\bibnamefont {Taniguchi}}, \bibinfo {author} {\bibfnamefont {K.}~\bibnamefont {Watanabe}}, \bibinfo {author} {\bibfnamefont {M.~A.}\ \bibnamefont {McGuire}}, \bibinfo {author} {\bibfnamefont {R.}~\bibnamefont {Stöhr}}, \bibinfo {author} {\bibfnamefont {D.}~\bibnamefont {Xiao}}, \bibinfo {author} {\bibfnamefont {T.}~\bibnamefont {Cao}}, \bibinfo {author} {\bibfnamefont {J.}~\bibnamefont {Wrachtrup}},\ and\ \bibinfo {author} {\bibfnamefont {X.}~\bibnamefont {Xu}},\ }\href {https://doi.org/10.1126/science.abj7478} {\bibfield  {journal} {\bibinfo  {journal} {Science}\ }\textbf {\bibinfo {volume} {374}},\ \bibinfo {pages} {1140} (\bibinfo {year}
  {2021})}\BibitemShut {NoStop}%
\bibitem [{\citenamefont {Dai}\ \emph {et~al.}(2019)\citenamefont {Dai}, \citenamefont {Liu},\ and\ \citenamefont {Zhang}}]{dai2019}%
  \BibitemOpen
  \bibfield  {author} {\bibinfo {author} {\bibfnamefont {Z.}~\bibnamefont {Dai}}, \bibinfo {author} {\bibfnamefont {L.}~\bibnamefont {Liu}},\ and\ \bibinfo {author} {\bibfnamefont {Z.}~\bibnamefont {Zhang}},\ }\href {https://doi.org/10.1002/adma.201805417} {\bibfield  {journal} {\bibinfo  {journal} {Advanced Materials}\ }\textbf {\bibinfo {volume} {31}},\ \bibinfo {pages} {1805417} (\bibinfo {year} {2019})}\BibitemShut {NoStop}%
\bibitem [{\citenamefont {Hsu}\ \emph {et~al.}(2020)\citenamefont {Hsu}, \citenamefont {Teague}, \citenamefont {Wang},\ and\ \citenamefont {Yeh}}]{hsu2020}%
  \BibitemOpen
  \bibfield  {author} {\bibinfo {author} {\bibfnamefont {C.-C.}\ \bibnamefont {Hsu}}, \bibinfo {author} {\bibfnamefont {M.~L.}\ \bibnamefont {Teague}}, \bibinfo {author} {\bibfnamefont {J.-Q.}\ \bibnamefont {Wang}},\ and\ \bibinfo {author} {\bibfnamefont {N.-C.}\ \bibnamefont {Yeh}},\ }\href {https://doi.org/10.1126/sciadv.aat9488} {\bibfield  {journal} {\bibinfo  {journal} {Science Advances}\ }\textbf {\bibinfo {volume} {6}},\ \bibinfo {pages} {eaat9488} (\bibinfo {year} {2020})}\BibitemShut {NoStop}%
\bibitem [{\citenamefont {Cenker}\ \emph {et~al.}(2022)\citenamefont {Cenker}, \citenamefont {Sivakumar}, \citenamefont {Xie}, \citenamefont {Miller}, \citenamefont {Thijssen}, \citenamefont {Liu}, \citenamefont {Dismukes}, \citenamefont {Fonseca}, \citenamefont {Anderson}, \citenamefont {Zhu}, \citenamefont {Roy}, \citenamefont {Xiao}, \citenamefont {Chu}, \citenamefont {Cao},\ and\ \citenamefont {Xu}}]{cenker2022}%
  \BibitemOpen
  \bibfield  {author} {\bibinfo {author} {\bibfnamefont {J.}~\bibnamefont {Cenker}}, \bibinfo {author} {\bibfnamefont {S.}~\bibnamefont {Sivakumar}}, \bibinfo {author} {\bibfnamefont {K.}~\bibnamefont {Xie}}, \bibinfo {author} {\bibfnamefont {A.}~\bibnamefont {Miller}}, \bibinfo {author} {\bibfnamefont {P.}~\bibnamefont {Thijssen}}, \bibinfo {author} {\bibfnamefont {Z.}~\bibnamefont {Liu}}, \bibinfo {author} {\bibfnamefont {A.}~\bibnamefont {Dismukes}}, \bibinfo {author} {\bibfnamefont {J.}~\bibnamefont {Fonseca}}, \bibinfo {author} {\bibfnamefont {E.}~\bibnamefont {Anderson}}, \bibinfo {author} {\bibfnamefont {X.}~\bibnamefont {Zhu}}, \bibinfo {author} {\bibfnamefont {X.}~\bibnamefont {Roy}}, \bibinfo {author} {\bibfnamefont {D.}~\bibnamefont {Xiao}}, \bibinfo {author} {\bibfnamefont {J.-H.}\ \bibnamefont {Chu}}, \bibinfo {author} {\bibfnamefont {T.}~\bibnamefont {Cao}},\ and\ \bibinfo {author} {\bibfnamefont {X.}~\bibnamefont {Xu}},\ }\href {https://doi.org/10.1038/s41565-021-01052-6} {\bibfield  {journal}
  {\bibinfo  {journal} {Nature Nanotechnology}\ }\textbf {\bibinfo {volume} {17}},\ \bibinfo {pages} {256} (\bibinfo {year} {2022})}\BibitemShut {NoStop}%
\bibitem [{\citenamefont {O’Hara}\ \emph {et~al.}(2018{\natexlab{b}})\citenamefont {O’Hara}, \citenamefont {Zhu}, \citenamefont {Trout}, \citenamefont {Ahmed}, \citenamefont {Luo}, \citenamefont {Lee}, \citenamefont {Brenner}, \citenamefont {Rajan}, \citenamefont {Gupta}, \citenamefont {McComb},\ and\ \citenamefont {Kawakami}}]{ohara2018}%
  \BibitemOpen
  \bibfield  {author} {\bibinfo {author} {\bibfnamefont {D.~J.}\ \bibnamefont {O’Hara}}, \bibinfo {author} {\bibfnamefont {T.}~\bibnamefont {Zhu}}, \bibinfo {author} {\bibfnamefont {A.~H.}\ \bibnamefont {Trout}}, \bibinfo {author} {\bibfnamefont {A.~S.}\ \bibnamefont {Ahmed}}, \bibinfo {author} {\bibfnamefont {Y.~K.}\ \bibnamefont {Luo}}, \bibinfo {author} {\bibfnamefont {C.~H.}\ \bibnamefont {Lee}}, \bibinfo {author} {\bibfnamefont {M.~R.}\ \bibnamefont {Brenner}}, \bibinfo {author} {\bibfnamefont {S.}~\bibnamefont {Rajan}}, \bibinfo {author} {\bibfnamefont {J.~A.}\ \bibnamefont {Gupta}}, \bibinfo {author} {\bibfnamefont {D.~W.}\ \bibnamefont {McComb}},\ and\ \bibinfo {author} {\bibfnamefont {R.~K.}\ \bibnamefont {Kawakami}},\ }\href {https://doi.org/10.1021/acs.nanolett.8b00683} {\bibfield  {journal} {\bibinfo  {journal} {Nano Lett.}\ }\textbf {\bibinfo {volume} {18}},\ \bibinfo {pages} {3125} (\bibinfo {year} {2018}{\natexlab{b}})}\BibitemShut {NoStop}%
\bibitem [{\citenamefont {Jin}\ \emph {et~al.}(2022)\citenamefont {Jin}, \citenamefont {Ji}, \citenamefont {Zhong}, \citenamefont {Jin}, \citenamefont {Hu}, \citenamefont {Zhang}, \citenamefont {Zhu}, \citenamefont {Huang}, \citenamefont {Li}, \citenamefont {Cai},\ and\ \citenamefont {Zhou}}]{jin2022}%
  \BibitemOpen
  \bibfield  {author} {\bibinfo {author} {\bibfnamefont {Z.}~\bibnamefont {Jin}}, \bibinfo {author} {\bibfnamefont {Z.}~\bibnamefont {Ji}}, \bibinfo {author} {\bibfnamefont {Y.}~\bibnamefont {Zhong}}, \bibinfo {author} {\bibfnamefont {Y.}~\bibnamefont {Jin}}, \bibinfo {author} {\bibfnamefont {X.}~\bibnamefont {Hu}}, \bibinfo {author} {\bibfnamefont {X.}~\bibnamefont {Zhang}}, \bibinfo {author} {\bibfnamefont {L.}~\bibnamefont {Zhu}}, \bibinfo {author} {\bibfnamefont {X.}~\bibnamefont {Huang}}, \bibinfo {author} {\bibfnamefont {T.}~\bibnamefont {Li}}, \bibinfo {author} {\bibfnamefont {X.}~\bibnamefont {Cai}},\ and\ \bibinfo {author} {\bibfnamefont {L.}~\bibnamefont {Zhou}},\ }\href {https://doi.org/10.1021/acsnano.1c11018} {\bibfield  {journal} {\bibinfo  {journal} {ACS Nano}\ }\textbf {\bibinfo {volume} {16}},\ \bibinfo {pages} {7572} (\bibinfo {year} {2022})}\BibitemShut {NoStop}%
\bibitem [{\citenamefont {Xie}\ \emph {et~al.}(2023)\citenamefont {Xie}, \citenamefont {Luo}, \citenamefont {Ye}, \citenamefont {Sun}, \citenamefont {Ye}, \citenamefont {Sung}, \citenamefont {Ge}, \citenamefont {Yan}, \citenamefont {Fu}, \citenamefont {Tian}, \citenamefont {Lei}, \citenamefont {Sun}, \citenamefont {Hovden}, \citenamefont {He},\ and\ \citenamefont {Zhao}}]{xie2023}%
  \BibitemOpen
  \bibfield  {author} {\bibinfo {author} {\bibfnamefont {H.}~\bibnamefont {Xie}}, \bibinfo {author} {\bibfnamefont {X.}~\bibnamefont {Luo}}, \bibinfo {author} {\bibfnamefont {Z.}~\bibnamefont {Ye}}, \bibinfo {author} {\bibfnamefont {Z.}~\bibnamefont {Sun}}, \bibinfo {author} {\bibfnamefont {G.}~\bibnamefont {Ye}}, \bibinfo {author} {\bibfnamefont {S.~H.}\ \bibnamefont {Sung}}, \bibinfo {author} {\bibfnamefont {H.}~\bibnamefont {Ge}}, \bibinfo {author} {\bibfnamefont {S.}~\bibnamefont {Yan}}, \bibinfo {author} {\bibfnamefont {Y.}~\bibnamefont {Fu}}, \bibinfo {author} {\bibfnamefont {S.}~\bibnamefont {Tian}}, \bibinfo {author} {\bibfnamefont {H.}~\bibnamefont {Lei}}, \bibinfo {author} {\bibfnamefont {K.}~\bibnamefont {Sun}}, \bibinfo {author} {\bibfnamefont {R.}~\bibnamefont {Hovden}}, \bibinfo {author} {\bibfnamefont {R.}~\bibnamefont {He}},\ and\ \bibinfo {author} {\bibfnamefont {L.}~\bibnamefont {Zhao}},\ }\href {https://doi.org/10.1038/s41567-023-02061-z} {\bibfield  {journal} {\bibinfo  {journal} {Nature
  Physics}\ }\textbf {\bibinfo {volume} {19}},\ \bibinfo {pages} {1150} (\bibinfo {year} {2023})}\BibitemShut {NoStop}%
\bibitem [{\citenamefont {Song}\ \emph {et~al.}(2022)\citenamefont {Song}, \citenamefont {Occhialini}, \citenamefont {Ergeçen}, \citenamefont {Ilyas}, \citenamefont {Amoroso}, \citenamefont {Barone}, \citenamefont {Kapeghian}, \citenamefont {Watanabe}, \citenamefont {Taniguchi}, \citenamefont {Botana}, \citenamefont {Picozzi}, \citenamefont {Gedik},\ and\ \citenamefont {Comin}}]{song2022}%
  \BibitemOpen
  \bibfield  {author} {\bibinfo {author} {\bibfnamefont {Q.}~\bibnamefont {Song}}, \bibinfo {author} {\bibfnamefont {C.~A.}\ \bibnamefont {Occhialini}}, \bibinfo {author} {\bibfnamefont {E.}~\bibnamefont {Ergeçen}}, \bibinfo {author} {\bibfnamefont {B.}~\bibnamefont {Ilyas}}, \bibinfo {author} {\bibfnamefont {D.}~\bibnamefont {Amoroso}}, \bibinfo {author} {\bibfnamefont {P.}~\bibnamefont {Barone}}, \bibinfo {author} {\bibfnamefont {J.}~\bibnamefont {Kapeghian}}, \bibinfo {author} {\bibfnamefont {K.}~\bibnamefont {Watanabe}}, \bibinfo {author} {\bibfnamefont {T.}~\bibnamefont {Taniguchi}}, \bibinfo {author} {\bibfnamefont {A.~S.}\ \bibnamefont {Botana}}, \bibinfo {author} {\bibfnamefont {S.}~\bibnamefont {Picozzi}}, \bibinfo {author} {\bibfnamefont {N.}~\bibnamefont {Gedik}},\ and\ \bibinfo {author} {\bibfnamefont {R.}~\bibnamefont {Comin}},\ }\href {https://doi.org/10.1038/s41586-021-04337-x} {\bibfield  {journal} {\bibinfo  {journal} {Nature}\ }\textbf {\bibinfo {volume} {602}},\ \bibinfo {pages} {601}
  (\bibinfo {year} {2022})}\BibitemShut {NoStop}%
\bibitem [{\citenamefont {Amini}\ \emph {et~al.}(2024)\citenamefont {Amini}, \citenamefont {Fumega}, \citenamefont {González-Herrero}, \citenamefont {Vaňo}, \citenamefont {Kezilebieke}, \citenamefont {Lado},\ and\ \citenamefont {Liljeroth}}]{amini2024}%
  \BibitemOpen
  \bibfield  {author} {\bibinfo {author} {\bibfnamefont {M.}~\bibnamefont {Amini}}, \bibinfo {author} {\bibfnamefont {A.~O.}\ \bibnamefont {Fumega}}, \bibinfo {author} {\bibfnamefont {H.}~\bibnamefont {González-Herrero}}, \bibinfo {author} {\bibfnamefont {V.}~\bibnamefont {Vaňo}}, \bibinfo {author} {\bibfnamefont {S.}~\bibnamefont {Kezilebieke}}, \bibinfo {author} {\bibfnamefont {J.~L.}\ \bibnamefont {Lado}},\ and\ \bibinfo {author} {\bibfnamefont {P.}~\bibnamefont {Liljeroth}},\ }\href {https://doi.org/10.1002/adma.202311342} {\bibfield  {journal} {\bibinfo  {journal} {Advanced Materials}\ }\textbf {\bibinfo {volume} {36}},\ \bibinfo {pages} {2311342} (\bibinfo {year} {2024})}\BibitemShut {NoStop}%
\bibitem [{\citenamefont {Lado}\ and\ \citenamefont {Fern{\'{a}}ndez-Rossier}(2017)}]{Lado2017}%
  \BibitemOpen
  \bibfield  {author} {\bibinfo {author} {\bibfnamefont {J.~L.}\ \bibnamefont {Lado}}\ and\ \bibinfo {author} {\bibfnamefont {J.}~\bibnamefont {Fern{\'{a}}ndez-Rossier}},\ }\href {https://doi.org/10.1088/2053-1583/aa75ed} {\bibfield  {journal} {\bibinfo  {journal} {2D Materials}\ }\textbf {\bibinfo {volume} {4}},\ \bibinfo {pages} {035002} (\bibinfo {year} {2017})}\BibitemShut {NoStop}%
\bibitem [{\citenamefont {Liu}\ \emph {et~al.}(2018)\citenamefont {Liu}, \citenamefont {Shi}, \citenamefont {Lu},\ and\ \citenamefont {Anantram}}]{liu2018}%
  \BibitemOpen
  \bibfield  {author} {\bibinfo {author} {\bibfnamefont {J.}~\bibnamefont {Liu}}, \bibinfo {author} {\bibfnamefont {M.}~\bibnamefont {Shi}}, \bibinfo {author} {\bibfnamefont {J.}~\bibnamefont {Lu}},\ and\ \bibinfo {author} {\bibfnamefont {M.~P.}\ \bibnamefont {Anantram}},\ }\href {https://doi.org/10.1103/PhysRevB.97.054416} {\bibfield  {journal} {\bibinfo  {journal} {Phys. Rev. B}\ }\textbf {\bibinfo {volume} {97}},\ \bibinfo {pages} {054416} (\bibinfo {year} {2018})}\BibitemShut {NoStop}%
\bibitem [{\citenamefont {Webster}\ and\ \citenamefont {Yan}(2018)}]{webster2018}%
  \BibitemOpen
  \bibfield  {author} {\bibinfo {author} {\bibfnamefont {L.}~\bibnamefont {Webster}}\ and\ \bibinfo {author} {\bibfnamefont {J.-A.}\ \bibnamefont {Yan}},\ }\href {https://doi.org/10.1103/PhysRevB.98.144411} {\bibfield  {journal} {\bibinfo  {journal} {Phys. Rev. B}\ }\textbf {\bibinfo {volume} {98}},\ \bibinfo {pages} {144411} (\bibinfo {year} {2018})}\BibitemShut {NoStop}%
\bibitem [{\citenamefont {Xu}\ \emph {et~al.}(2018)\citenamefont {Xu}, \citenamefont {Feng}, \citenamefont {Xiang},\ and\ \citenamefont {Bellaiche}}]{Kitaev2018}%
  \BibitemOpen
  \bibfield  {author} {\bibinfo {author} {\bibfnamefont {C.}~\bibnamefont {Xu}}, \bibinfo {author} {\bibfnamefont {J.}~\bibnamefont {Feng}}, \bibinfo {author} {\bibfnamefont {H.}~\bibnamefont {Xiang}},\ and\ \bibinfo {author} {\bibfnamefont {L.}~\bibnamefont {Bellaiche}},\ }\href {https://doi.org/10.1038/s41524-018-0115-6} {\bibfield  {journal} {\bibinfo  {journal} {npj Comput. Mater.}\ }\textbf {\bibinfo {volume} {4}},\ \bibinfo {pages} {0} (\bibinfo {year} {2018})}\BibitemShut {NoStop}%
\bibitem [{\citenamefont {Torelli}\ and\ \citenamefont {Olsen}(2018)}]{torelli2018}%
  \BibitemOpen
  \bibfield  {author} {\bibinfo {author} {\bibfnamefont {D.}~\bibnamefont {Torelli}}\ and\ \bibinfo {author} {\bibfnamefont {T.}~\bibnamefont {Olsen}},\ }\href {https://doi.org/10.1088/2053-1583/aaf06d} {\bibfield  {journal} {\bibinfo  {journal} {2D Materials}\ }\textbf {\bibinfo {volume} {6}},\ \bibinfo {pages} {015028} (\bibinfo {year} {2018})}\BibitemShut {NoStop}%
\bibitem [{\citenamefont {Jang}\ \emph {et~al.}(2019)\citenamefont {Jang}, \citenamefont {Jeong}, \citenamefont {Yoon}, \citenamefont {Ryee},\ and\ \citenamefont {Han}}]{woojang2019}%
  \BibitemOpen
  \bibfield  {author} {\bibinfo {author} {\bibfnamefont {S.~W.}\ \bibnamefont {Jang}}, \bibinfo {author} {\bibfnamefont {M.~Y.}\ \bibnamefont {Jeong}}, \bibinfo {author} {\bibfnamefont {H.}~\bibnamefont {Yoon}}, \bibinfo {author} {\bibfnamefont {S.}~\bibnamefont {Ryee}},\ and\ \bibinfo {author} {\bibfnamefont {M.~J.}\ \bibnamefont {Han}},\ }\href {https://doi.org/10.1103/PhysRevMaterials.3.031001} {\bibfield  {journal} {\bibinfo  {journal} {Phys. Rev. Mater.}\ }\textbf {\bibinfo {volume} {3}},\ \bibinfo {pages} {031001} (\bibinfo {year} {2019})}\BibitemShut {NoStop}%
\bibitem [{\citenamefont {Kim}\ \emph {et~al.}(2019)\citenamefont {Kim}, \citenamefont {Kim}, \citenamefont {Ko}, \citenamefont {Seo}, \citenamefont {Kim}, \citenamefont {Jang}, \citenamefont {Kim}, \citenamefont {Kim}, \citenamefont {Cheong},\ and\ \citenamefont {Park}}]{kim2019}%
  \BibitemOpen
  \bibfield  {author} {\bibinfo {author} {\bibfnamefont {D.-H.}\ \bibnamefont {Kim}}, \bibinfo {author} {\bibfnamefont {K.}~\bibnamefont {Kim}}, \bibinfo {author} {\bibfnamefont {K.-T.}\ \bibnamefont {Ko}}, \bibinfo {author} {\bibfnamefont {J.}~\bibnamefont {Seo}}, \bibinfo {author} {\bibfnamefont {J.~S.}\ \bibnamefont {Kim}}, \bibinfo {author} {\bibfnamefont {T.-H.}\ \bibnamefont {Jang}}, \bibinfo {author} {\bibfnamefont {Y.}~\bibnamefont {Kim}}, \bibinfo {author} {\bibfnamefont {J.-Y.}\ \bibnamefont {Kim}}, \bibinfo {author} {\bibfnamefont {S.-W.}\ \bibnamefont {Cheong}},\ and\ \bibinfo {author} {\bibfnamefont {J.-H.}\ \bibnamefont {Park}},\ }\href {https://doi.org/10.1103/PhysRevLett.122.207201} {\bibfield  {journal} {\bibinfo  {journal} {Phys. Rev. Lett.}\ }\textbf {\bibinfo {volume} {122}},\ \bibinfo {pages} {207201} (\bibinfo {year} {2019})}\BibitemShut {NoStop}%
\bibitem [{\citenamefont {Jiang}\ \emph {et~al.}(2019)\citenamefont {Jiang}, \citenamefont {Wang}, \citenamefont {Chen}, \citenamefont {Zhong}, \citenamefont {Yuan}, \citenamefont {Lu},\ and\ \citenamefont {Ji}}]{jiang2019}%
  \BibitemOpen
  \bibfield  {author} {\bibinfo {author} {\bibfnamefont {P.}~\bibnamefont {Jiang}}, \bibinfo {author} {\bibfnamefont {C.}~\bibnamefont {Wang}}, \bibinfo {author} {\bibfnamefont {D.}~\bibnamefont {Chen}}, \bibinfo {author} {\bibfnamefont {Z.}~\bibnamefont {Zhong}}, \bibinfo {author} {\bibfnamefont {Z.}~\bibnamefont {Yuan}}, \bibinfo {author} {\bibfnamefont {Z.-Y.}\ \bibnamefont {Lu}},\ and\ \bibinfo {author} {\bibfnamefont {W.}~\bibnamefont {Ji}},\ }\href {https://doi.org/10.1103/PhysRevB.99.144401} {\bibfield  {journal} {\bibinfo  {journal} {Phys. Rev. B}\ }\textbf {\bibinfo {volume} {99}},\ \bibinfo {pages} {144401} (\bibinfo {year} {2019})}\BibitemShut {NoStop}%
\bibitem [{\citenamefont {Soriano}\ \emph {et~al.}(2019)\citenamefont {Soriano}, \citenamefont {Cardoso},\ and\ \citenamefont {Fernandez-Rosier}}]{soriano2019}%
  \BibitemOpen
  \bibfield  {author} {\bibinfo {author} {\bibfnamefont {D.}~\bibnamefont {Soriano}}, \bibinfo {author} {\bibfnamefont {C.}~\bibnamefont {Cardoso}},\ and\ \bibinfo {author} {\bibfnamefont {J.}~\bibnamefont {Fernandez-Rosier}},\ }\href {https://doi.org/10.1016/j.ssc.2019.113662} {\bibfield  {journal} {\bibinfo  {journal} {Solid State Communications}\ }\textbf {\bibinfo {volume} {299}},\ \bibinfo {pages} {113662} (\bibinfo {year} {2019})}\BibitemShut {NoStop}%
\bibitem [{\citenamefont {Zhang}\ \emph {et~al.}(2019)\citenamefont {Zhang}, \citenamefont {Kong}, \citenamefont {Pang}, \citenamefont {Hu}, \citenamefont {Gong}, \citenamefont {Shi},\ and\ \citenamefont {Tang}}]{zhang2019}%
  \BibitemOpen
  \bibfield  {author} {\bibinfo {author} {\bibfnamefont {F.}~\bibnamefont {Zhang}}, \bibinfo {author} {\bibfnamefont {Y.-C.}\ \bibnamefont {Kong}}, \bibinfo {author} {\bibfnamefont {R.}~\bibnamefont {Pang}}, \bibinfo {author} {\bibfnamefont {L.}~\bibnamefont {Hu}}, \bibinfo {author} {\bibfnamefont {P.-L.}\ \bibnamefont {Gong}}, \bibinfo {author} {\bibfnamefont {X.-Q.}\ \bibnamefont {Shi}},\ and\ \bibinfo {author} {\bibfnamefont {Z.-K.}\ \bibnamefont {Tang}},\ }\href {https://doi.org/10.1088/1367-2630/ab1ee4} {\bibfield  {journal} {\bibinfo  {journal} {New J. Phys.}\ }\textbf {\bibinfo {volume} {21}},\ \bibinfo {pages} {053033} (\bibinfo {year} {2019})}\BibitemShut {NoStop}%
\bibitem [{\citenamefont {Kashin}\ \emph {et~al.}(2020)\citenamefont {Kashin}, \citenamefont {Mazurenko}, \citenamefont {Katsnelson},\ and\ \citenamefont {Rudenko}}]{Kashin2020}%
  \BibitemOpen
  \bibfield  {author} {\bibinfo {author} {\bibfnamefont {I.~V.}\ \bibnamefont {Kashin}}, \bibinfo {author} {\bibfnamefont {V.~V.}\ \bibnamefont {Mazurenko}}, \bibinfo {author} {\bibfnamefont {M.~I.}\ \bibnamefont {Katsnelson}},\ and\ \bibinfo {author} {\bibfnamefont {A.~N.}\ \bibnamefont {Rudenko}},\ }\href {https://doi.org/10.1088/2053-1583/ab72d8} {\bibfield  {journal} {\bibinfo  {journal} {2D Materials}\ }\textbf {\bibinfo {volume} {7}},\ \bibinfo {pages} {025036} (\bibinfo {year} {2020})}\BibitemShut {NoStop}%
\bibitem [{\citenamefont {Xu}\ \emph {et~al.}(2020)\citenamefont {Xu}, \citenamefont {Chen}, \citenamefont {Tan}, \citenamefont {Yang}, \citenamefont {Xiang},\ and\ \citenamefont {Bellaiche}}]{cxu2020}%
  \BibitemOpen
  \bibfield  {author} {\bibinfo {author} {\bibfnamefont {C.}~\bibnamefont {Xu}}, \bibinfo {author} {\bibfnamefont {P.}~\bibnamefont {Chen}}, \bibinfo {author} {\bibfnamefont {H.}~\bibnamefont {Tan}}, \bibinfo {author} {\bibfnamefont {Y.}~\bibnamefont {Yang}}, \bibinfo {author} {\bibfnamefont {H.}~\bibnamefont {Xiang}},\ and\ \bibinfo {author} {\bibfnamefont {L.}~\bibnamefont {Bellaiche}},\ }\href {https://doi.org/10.1103/PhysRevLett.125.037203} {\bibfield  {journal} {\bibinfo  {journal} {Phys. Rev. Lett.}\ }\textbf {\bibinfo {volume} {125}},\ \bibinfo {pages} {037203} (\bibinfo {year} {2020})}\BibitemShut {NoStop}%
\bibitem [{\citenamefont {Kvashnin}\ \emph {et~al.}(2020)\citenamefont {Kvashnin}, \citenamefont {Bergman}, \citenamefont {Lichtenstein},\ and\ \citenamefont {Katsnelson}}]{kvashnin2020}%
  \BibitemOpen
  \bibfield  {author} {\bibinfo {author} {\bibfnamefont {Y.~O.}\ \bibnamefont {Kvashnin}}, \bibinfo {author} {\bibfnamefont {A.}~\bibnamefont {Bergman}}, \bibinfo {author} {\bibfnamefont {A.~I.}\ \bibnamefont {Lichtenstein}},\ and\ \bibinfo {author} {\bibfnamefont {M.~I.}\ \bibnamefont {Katsnelson}},\ }\href {https://doi.org/10.1103/PhysRevB.102.115162} {\bibfield  {journal} {\bibinfo  {journal} {Phys. Rev. B}\ }\textbf {\bibinfo {volume} {102}},\ \bibinfo {pages} {115162} (\bibinfo {year} {2020})}\BibitemShut {NoStop}%
\bibitem [{\citenamefont {Pizzochero}\ \emph {et~al.}(2020)\citenamefont {Pizzochero}, \citenamefont {Yadav},\ and\ \citenamefont {Yazyen}}]{pizzochero2020}%
  \BibitemOpen
  \bibfield  {author} {\bibinfo {author} {\bibfnamefont {M.}~\bibnamefont {Pizzochero}}, \bibinfo {author} {\bibfnamefont {R.}~\bibnamefont {Yadav}},\ and\ \bibinfo {author} {\bibfnamefont {O.~V.}\ \bibnamefont {Yazyen}},\ }\href {https://doi.org/10.1088/2053-1583/ab7cab} {\bibfield  {journal} {\bibinfo  {journal} {2D Materials}\ }\textbf {\bibinfo {volume} {7}},\ \bibinfo {pages} {035005} (\bibinfo {year} {2020})}\BibitemShut {NoStop}%
\bibitem [{\citenamefont {Bacaksiz}\ \emph {et~al.}(2021)\citenamefont {Bacaksiz}, \citenamefont {\ifmmode~\check{S}\else \v{S}\fi{}abani}, \citenamefont {Menezes},\ and\ \citenamefont {Milo\ifmmode \check{s}\else \v{s}\fi{}evi\ifmmode~\acute{c}\else \'{c}\fi{}}}]{bacaksiz2021}%
  \BibitemOpen
  \bibfield  {author} {\bibinfo {author} {\bibfnamefont {C.}~\bibnamefont {Bacaksiz}}, \bibinfo {author} {\bibfnamefont {D.}~\bibnamefont {\ifmmode~\check{S}\else \v{S}\fi{}abani}}, \bibinfo {author} {\bibfnamefont {R.~M.}\ \bibnamefont {Menezes}},\ and\ \bibinfo {author} {\bibfnamefont {M.~V.}\ \bibnamefont {Milo\ifmmode \check{s}\else \v{s}\fi{}evi\ifmmode~\acute{c}\else \'{c}\fi{}}},\ }\href {https://doi.org/10.1103/PhysRevB.103.125418} {\bibfield  {journal} {\bibinfo  {journal} {Phys. Rev. B}\ }\textbf {\bibinfo {volume} {103}},\ \bibinfo {pages} {125418} (\bibinfo {year} {2021})}\BibitemShut {NoStop}%
\bibitem [{\citenamefont {Ni}\ \emph {et~al.}(2021)\citenamefont {Ni}, \citenamefont {Li}, \citenamefont {Amoroso}, \citenamefont {He}, \citenamefont {Feng}, \citenamefont {Kan}, \citenamefont {Picozzi},\ and\ \citenamefont {Xiang}}]{ni2021}%
  \BibitemOpen
  \bibfield  {author} {\bibinfo {author} {\bibfnamefont {J.~Y.}\ \bibnamefont {Ni}}, \bibinfo {author} {\bibfnamefont {X.~Y.}\ \bibnamefont {Li}}, \bibinfo {author} {\bibfnamefont {D.}~\bibnamefont {Amoroso}}, \bibinfo {author} {\bibfnamefont {X.}~\bibnamefont {He}}, \bibinfo {author} {\bibfnamefont {J.~S.}\ \bibnamefont {Feng}}, \bibinfo {author} {\bibfnamefont {E.~J.}\ \bibnamefont {Kan}}, \bibinfo {author} {\bibfnamefont {S.}~\bibnamefont {Picozzi}},\ and\ \bibinfo {author} {\bibfnamefont {H.~J.}\ \bibnamefont {Xiang}},\ }\href {https://doi.org/10.1103/PhysRevLett.127.247204} {\bibfield  {journal} {\bibinfo  {journal} {Phys. Rev. Lett.}\ }\textbf {\bibinfo {volume} {127}},\ \bibinfo {pages} {247204} (\bibinfo {year} {2021})}\BibitemShut {NoStop}%
\bibitem [{\citenamefont {Xu}\ \emph {et~al.}(2022)\citenamefont {Xu}, \citenamefont {Li}, \citenamefont {Chen}, \citenamefont {Zhang}, \citenamefont {Xiang},\ and\ \citenamefont {Bellaiche}}]{cxu2022}%
  \BibitemOpen
  \bibfield  {author} {\bibinfo {author} {\bibfnamefont {C.}~\bibnamefont {Xu}}, \bibinfo {author} {\bibfnamefont {X.}~\bibnamefont {Li}}, \bibinfo {author} {\bibfnamefont {P.}~\bibnamefont {Chen}}, \bibinfo {author} {\bibfnamefont {Y.}~\bibnamefont {Zhang}}, \bibinfo {author} {\bibfnamefont {H.}~\bibnamefont {Xiang}},\ and\ \bibinfo {author} {\bibfnamefont {L.}~\bibnamefont {Bellaiche}},\ }\href {https://doi.org/10.1002/adma.202107779} {\bibfield  {journal} {\bibinfo  {journal} {Advanced Materials}\ }\textbf {\bibinfo {volume} {34}},\ \bibinfo {pages} {2107779} (\bibinfo {year} {2022})}\BibitemShut {NoStop}%
\bibitem [{\citenamefont {Fumega}\ and\ \citenamefont {L.}(2022)}]{fumega2022}%
  \BibitemOpen
  \bibfield  {author} {\bibinfo {author} {\bibfnamefont {A.~O.}\ \bibnamefont {Fumega}}\ and\ \bibinfo {author} {\bibfnamefont {L.~J.}\ \bibnamefont {L.}},\ }\href {https://doi.org/10.1088/2053-1583/ac4e9d} {\bibfield  {journal} {\bibinfo  {journal} {2D Materials}\ }\textbf {\bibinfo {volume} {9}},\ \bibinfo {pages} {025010} (\bibinfo {year} {2022})}\BibitemShut {NoStop}%
\bibitem [{\citenamefont {Edstr\"om}\ \emph {et~al.}(2022)\citenamefont {Edstr\"om}, \citenamefont {Amoroso}, \citenamefont {Picozzi}, \citenamefont {Barone},\ and\ \citenamefont {Stengel}}]{edstrom2022}%
  \BibitemOpen
  \bibfield  {author} {\bibinfo {author} {\bibfnamefont {A.}~\bibnamefont {Edstr\"om}}, \bibinfo {author} {\bibfnamefont {D.}~\bibnamefont {Amoroso}}, \bibinfo {author} {\bibfnamefont {S.}~\bibnamefont {Picozzi}}, \bibinfo {author} {\bibfnamefont {P.}~\bibnamefont {Barone}},\ and\ \bibinfo {author} {\bibfnamefont {M.}~\bibnamefont {Stengel}},\ }\href {https://doi.org/10.1103/PhysRevLett.128.177202} {\bibfield  {journal} {\bibinfo  {journal} {Phys. Rev. Lett.}\ }\textbf {\bibinfo {volume} {128}},\ \bibinfo {pages} {177202} (\bibinfo {year} {2022})}\BibitemShut {NoStop}%
\bibitem [{\citenamefont {Riedl}\ \emph {et~al.}(2022)\citenamefont {Riedl}, \citenamefont {Amoroso}, \citenamefont {Backes}, \citenamefont {Razpopov}, \citenamefont {Nguyen}, \citenamefont {Yamauchi}, \citenamefont {Barone}, \citenamefont {Winter}, \citenamefont {Picozzi},\ and\ \citenamefont {Valent\'{\i}}}]{riedl2022}%
  \BibitemOpen
  \bibfield  {author} {\bibinfo {author} {\bibfnamefont {K.}~\bibnamefont {Riedl}}, \bibinfo {author} {\bibfnamefont {D.}~\bibnamefont {Amoroso}}, \bibinfo {author} {\bibfnamefont {S.}~\bibnamefont {Backes}}, \bibinfo {author} {\bibfnamefont {A.}~\bibnamefont {Razpopov}}, \bibinfo {author} {\bibfnamefont {T.~P.~T.}\ \bibnamefont {Nguyen}}, \bibinfo {author} {\bibfnamefont {K.}~\bibnamefont {Yamauchi}}, \bibinfo {author} {\bibfnamefont {P.}~\bibnamefont {Barone}}, \bibinfo {author} {\bibfnamefont {S.~M.}\ \bibnamefont {Winter}}, \bibinfo {author} {\bibfnamefont {S.}~\bibnamefont {Picozzi}},\ and\ \bibinfo {author} {\bibfnamefont {R.}~\bibnamefont {Valent\'{\i}}},\ }\href {https://doi.org/10.1103/PhysRevB.106.035156} {\bibfield  {journal} {\bibinfo  {journal} {Phys. Rev. B}\ }\textbf {\bibinfo {volume} {106}},\ \bibinfo {pages} {035156} (\bibinfo {year} {2022})}\BibitemShut {NoStop}%
\bibitem [{\citenamefont {Bo}\ \emph {et~al.}(2023{\natexlab{a}})\citenamefont {Bo}, \citenamefont {Li}, \citenamefont {Xu}, \citenamefont {Wan},\ and\ \citenamefont {Pu}}]{bo2023}%
  \BibitemOpen
  \bibfield  {author} {\bibinfo {author} {\bibfnamefont {X.}~\bibnamefont {Bo}}, \bibinfo {author} {\bibfnamefont {F.}~\bibnamefont {Li}}, \bibinfo {author} {\bibfnamefont {X.}~\bibnamefont {Xu}}, \bibinfo {author} {\bibfnamefont {X.}~\bibnamefont {Wan}},\ and\ \bibinfo {author} {\bibfnamefont {Y.}~\bibnamefont {Pu}},\ }\href {https://doi.org/10.1088/1367-2630/acb3ee} {\bibfield  {journal} {\bibinfo  {journal} {New J. Phys.}\ }\textbf {\bibinfo {volume} {25}},\ \bibinfo {pages} {013026} (\bibinfo {year} {2023}{\natexlab{a}})}\BibitemShut {NoStop}%
\bibitem [{\citenamefont {Bo}\ \emph {et~al.}(2023{\natexlab{b}})\citenamefont {Bo}, \citenamefont {Li}, \citenamefont {Yin}, \citenamefont {Chen}, \citenamefont {Wan},\ and\ \citenamefont {Pu}}]{bo2023b}%
  \BibitemOpen
  \bibfield  {author} {\bibinfo {author} {\bibfnamefont {X.}~\bibnamefont {Bo}}, \bibinfo {author} {\bibfnamefont {F.}~\bibnamefont {Li}}, \bibinfo {author} {\bibfnamefont {X.}~\bibnamefont {Yin}}, \bibinfo {author} {\bibfnamefont {Y.}~\bibnamefont {Chen}}, \bibinfo {author} {\bibfnamefont {X.}~\bibnamefont {Wan}},\ and\ \bibinfo {author} {\bibfnamefont {Y.}~\bibnamefont {Pu}},\ }\href {https://doi.org/10.1103/PhysRevB.108.024405} {\bibfield  {journal} {\bibinfo  {journal} {Phys. Rev. B}\ }\textbf {\bibinfo {volume} {108}},\ \bibinfo {pages} {024405} (\bibinfo {year} {2023}{\natexlab{b}})}\BibitemShut {NoStop}%
\bibitem [{\citenamefont {Wang}\ \emph {et~al.}(2023)\citenamefont {Wang}, \citenamefont {Yang}, \citenamefont {Ma}, \citenamefont {Liu}, \citenamefont {Lu}, \citenamefont {Zhou},\ and\ \citenamefont {Wu}}]{wang2023}%
  \BibitemOpen
  \bibfield  {author} {\bibinfo {author} {\bibfnamefont {G.}~\bibnamefont {Wang}}, \bibinfo {author} {\bibfnamefont {K.}~\bibnamefont {Yang}}, \bibinfo {author} {\bibfnamefont {Y.}~\bibnamefont {Ma}}, \bibinfo {author} {\bibfnamefont {L.}~\bibnamefont {Liu}}, \bibinfo {author} {\bibfnamefont {D.}~\bibnamefont {Lu}}, \bibinfo {author} {\bibfnamefont {Y.}~\bibnamefont {Zhou}},\ and\ \bibinfo {author} {\bibfnamefont {H.}~\bibnamefont {Wu}},\ }\href {https://doi.org/10.1088/0256-307X/40/7/077301} {\bibfield  {journal} {\bibinfo  {journal} {Chinese Phys. Lett.}\ }\textbf {\bibinfo {volume} {40}},\ \bibinfo {pages} {077301} (\bibinfo {year} {2023})}\BibitemShut {NoStop}%
\bibitem [{\citenamefont {Bo}\ \emph {et~al.}(2024)\citenamefont {Bo}, \citenamefont {Fu}, \citenamefont {Wan}, \citenamefont {Li},\ and\ \citenamefont {Pu}}]{bo2024}%
  \BibitemOpen
  \bibfield  {author} {\bibinfo {author} {\bibfnamefont {X.}~\bibnamefont {Bo}}, \bibinfo {author} {\bibfnamefont {L.}~\bibnamefont {Fu}}, \bibinfo {author} {\bibfnamefont {X.}~\bibnamefont {Wan}}, \bibinfo {author} {\bibfnamefont {S.}~\bibnamefont {Li}},\ and\ \bibinfo {author} {\bibfnamefont {Y.}~\bibnamefont {Pu}},\ }\href {https://doi.org/10.1103/PhysRevB.109.014405} {\bibfield  {journal} {\bibinfo  {journal} {Phys. Rev. B}\ }\textbf {\bibinfo {volume} {109}},\ \bibinfo {pages} {014405} (\bibinfo {year} {2024})}\BibitemShut {NoStop}%
\bibitem [{\citenamefont {Yorulmaz}\ \emph {et~al.}(2024)\citenamefont {Yorulmaz}, \citenamefont {Šabani}, \citenamefont {Sevik},\ and\ \citenamefont {Milošević}}]{yorulmaz2024}%
  \BibitemOpen
  \bibfield  {author} {\bibinfo {author} {\bibfnamefont {U.}~\bibnamefont {Yorulmaz}}, \bibinfo {author} {\bibfnamefont {D.}~\bibnamefont {Šabani}}, \bibinfo {author} {\bibfnamefont {C.}~\bibnamefont {Sevik}},\ and\ \bibinfo {author} {\bibfnamefont {M.~V.}\ \bibnamefont {Milošević}},\ }\href {https://doi.org/10.1088/2053-1583/ad3e08} {\bibfield  {journal} {\bibinfo  {journal} {2D Materials}\ }\textbf {\bibinfo {volume} {11}},\ \bibinfo {pages} {035013} (\bibinfo {year} {2024})}\BibitemShut {NoStop}%
\bibitem [{\citenamefont {Liu}\ \emph {et~al.}(2024{\natexlab{a}})\citenamefont {Liu}, \citenamefont {Wang}, \citenamefont {Yan}, \citenamefont {Xu}, \citenamefont {Hu}, \citenamefont {Zhang},\ and\ \citenamefont {Ji}}]{liu2024}%
  \BibitemOpen
  \bibfield  {author} {\bibinfo {author} {\bibfnamefont {N.}~\bibnamefont {Liu}}, \bibinfo {author} {\bibfnamefont {C.}~\bibnamefont {Wang}}, \bibinfo {author} {\bibfnamefont {C.}~\bibnamefont {Yan}}, \bibinfo {author} {\bibfnamefont {C.}~\bibnamefont {Xu}}, \bibinfo {author} {\bibfnamefont {J.}~\bibnamefont {Hu}}, \bibinfo {author} {\bibfnamefont {Y.}~\bibnamefont {Zhang}},\ and\ \bibinfo {author} {\bibfnamefont {W.}~\bibnamefont {Ji}},\ }\href {https://doi.org/10.1103/PhysRevB.109.195422} {\bibfield  {journal} {\bibinfo  {journal} {Phys. Rev. B}\ }\textbf {\bibinfo {volume} {109}},\ \bibinfo {pages} {195422} (\bibinfo {year} {2024}{\natexlab{a}})}\BibitemShut {NoStop}%
\bibitem [{\citenamefont {Anderson}(1963)}]{anderson2}%
  \BibitemOpen
  \bibfield  {author} {\bibinfo {author} {\bibfnamefont {P.~W.}\ \bibnamefont {Anderson}},\ }\href {https://doi.org/https://doi.org/10.1016/S0081-1947(08)60260-X} {\bibfield  {journal} {\bibinfo  {journal} {Solid State Physics}\ }\textbf {\bibinfo {volume} {14}},\ \bibinfo {pages} {99} (\bibinfo {year} {1963})}\BibitemShut {NoStop}%
\bibitem [{\citenamefont {Liechtenstein}\ \emph {et~al.}(1987)\citenamefont {Liechtenstein}, \citenamefont {Katsnelson}, \citenamefont {Antropov},\ and\ \citenamefont {Gubanov}}]{LKAG1987}%
  \BibitemOpen
  \bibfield  {author} {\bibinfo {author} {\bibfnamefont {A.}~\bibnamefont {Liechtenstein}}, \bibinfo {author} {\bibfnamefont {M.}~\bibnamefont {Katsnelson}}, \bibinfo {author} {\bibfnamefont {V.}~\bibnamefont {Antropov}},\ and\ \bibinfo {author} {\bibfnamefont {V.}~\bibnamefont {Gubanov}},\ }\href {https://doi.org/https://doi.org/10.1016/0304-8853(87)90721-9} {\bibfield  {journal} {\bibinfo  {journal} {Journal of Magnetism and Magnetic Materials}\ }\textbf {\bibinfo {volume} {67}},\ \bibinfo {pages} {65} (\bibinfo {year} {1987})}\BibitemShut {NoStop}%
\bibitem [{\citenamefont {He}\ \emph {et~al.}(2021)\citenamefont {He}, \citenamefont {Helbig}, \citenamefont {Verstraete},\ and\ \citenamefont {Bousquet}}]{TB2J2021}%
  \BibitemOpen
  \bibfield  {author} {\bibinfo {author} {\bibfnamefont {X.}~\bibnamefont {He}}, \bibinfo {author} {\bibfnamefont {N.}~\bibnamefont {Helbig}}, \bibinfo {author} {\bibfnamefont {M.~J.}\ \bibnamefont {Verstraete}},\ and\ \bibinfo {author} {\bibfnamefont {E.}~\bibnamefont {Bousquet}},\ }\href {https://doi.org/https://doi.org/10.1016/j.cpc.2021.107938} {\bibfield  {journal} {\bibinfo  {journal} {Computer Physics Communications}\ }\textbf {\bibinfo {volume} {264}},\ \bibinfo {pages} {107938} (\bibinfo {year} {2021})}\BibitemShut {NoStop}%
\bibitem [{\citenamefont {Kresse}\ and\ \citenamefont {Hafner}(1993)}]{vasp1}%
  \BibitemOpen
  \bibfield  {author} {\bibinfo {author} {\bibfnamefont {G.}~\bibnamefont {Kresse}}\ and\ \bibinfo {author} {\bibfnamefont {J.}~\bibnamefont {Hafner}},\ }\href {https://doi.org/10.1103/PhysRevB.47.558} {\bibfield  {journal} {\bibinfo  {journal} {Phys. Rev. B}\ }\textbf {\bibinfo {volume} {47}},\ \bibinfo {pages} {558} (\bibinfo {year} {1993})}\BibitemShut {NoStop}%
\bibitem [{\citenamefont {Kresse}\ and\ \citenamefont {Furthmüller}(1996)}]{vasp2}%
  \BibitemOpen
  \bibfield  {author} {\bibinfo {author} {\bibfnamefont {G.}~\bibnamefont {Kresse}}\ and\ \bibinfo {author} {\bibfnamefont {J.}~\bibnamefont {Furthmüller}},\ }\href {https://doi.org/https://doi.org/10.1016/0927-0256(96)00008-0} {\bibfield  {journal} {\bibinfo  {journal} {Computational Materials Science}\ }\textbf {\bibinfo {volume} {6}},\ \bibinfo {pages} {15} (\bibinfo {year} {1996})}\BibitemShut {NoStop}%
\bibitem [{\citenamefont {Kresse}\ and\ \citenamefont {Furthm\"uller}(1996)}]{vasp3}%
  \BibitemOpen
  \bibfield  {author} {\bibinfo {author} {\bibfnamefont {G.}~\bibnamefont {Kresse}}\ and\ \bibinfo {author} {\bibfnamefont {J.}~\bibnamefont {Furthm\"uller}},\ }\href {https://doi.org/10.1103/PhysRevB.54.11169} {\bibfield  {journal} {\bibinfo  {journal} {Phys. Rev. B}\ }\textbf {\bibinfo {volume} {54}},\ \bibinfo {pages} {11169} (\bibinfo {year} {1996})}\BibitemShut {NoStop}%
\bibitem [{\citenamefont {Pizzi}\ \emph {et~al.}(2020)\citenamefont {Pizzi}, \citenamefont {Vitale}, \citenamefont {Arita}, \citenamefont {Blügel}, \citenamefont {Freimuth}, \citenamefont {Géranton}, \citenamefont {Gibertini}, \citenamefont {Gresch}, \citenamefont {Johnson}, \citenamefont {Koretsune}, \citenamefont {Ibañez-Azpiroz}, \citenamefont {Lee}, \citenamefont {Lihm}, \citenamefont {Marchand}, \citenamefont {Marrazzo}, \citenamefont {Mokrousov}, \citenamefont {Mustafa}, \citenamefont {Nohara}, \citenamefont {Nomura}, \citenamefont {Paulatto}, \citenamefont {Poncé}, \citenamefont {Ponweiser}, \citenamefont {Qiao}, \citenamefont {Thöle}, \citenamefont {Tsirkin}, \citenamefont {Wierzbowska}, \citenamefont {Marzari}, \citenamefont {Vanderbilt}, \citenamefont {Souza}, \citenamefont {Mostofi},\ and\ \citenamefont {Yates}}]{wannier90}%
  \BibitemOpen
  \bibfield  {author} {\bibinfo {author} {\bibfnamefont {G.}~\bibnamefont {Pizzi}}, \bibinfo {author} {\bibfnamefont {V.}~\bibnamefont {Vitale}}, \bibinfo {author} {\bibfnamefont {R.}~\bibnamefont {Arita}}, \bibinfo {author} {\bibfnamefont {S.}~\bibnamefont {Blügel}}, \bibinfo {author} {\bibfnamefont {F.}~\bibnamefont {Freimuth}}, \bibinfo {author} {\bibfnamefont {G.}~\bibnamefont {Géranton}}, \bibinfo {author} {\bibfnamefont {M.}~\bibnamefont {Gibertini}}, \bibinfo {author} {\bibfnamefont {D.}~\bibnamefont {Gresch}}, \bibinfo {author} {\bibfnamefont {C.}~\bibnamefont {Johnson}}, \bibinfo {author} {\bibfnamefont {T.}~\bibnamefont {Koretsune}}, \bibinfo {author} {\bibfnamefont {J.}~\bibnamefont {Ibañez-Azpiroz}}, \bibinfo {author} {\bibfnamefont {H.}~\bibnamefont {Lee}}, \bibinfo {author} {\bibfnamefont {J.-M.}\ \bibnamefont {Lihm}}, \bibinfo {author} {\bibfnamefont {D.}~\bibnamefont {Marchand}}, \bibinfo {author} {\bibfnamefont {A.}~\bibnamefont {Marrazzo}}, \bibinfo {author} {\bibfnamefont
  {Y.}~\bibnamefont {Mokrousov}}, \bibinfo {author} {\bibfnamefont {J.~I.}\ \bibnamefont {Mustafa}}, \bibinfo {author} {\bibfnamefont {Y.}~\bibnamefont {Nohara}}, \bibinfo {author} {\bibfnamefont {Y.}~\bibnamefont {Nomura}}, \bibinfo {author} {\bibfnamefont {L.}~\bibnamefont {Paulatto}}, \bibinfo {author} {\bibfnamefont {S.}~\bibnamefont {Poncé}}, \bibinfo {author} {\bibfnamefont {T.}~\bibnamefont {Ponweiser}}, \bibinfo {author} {\bibfnamefont {J.}~\bibnamefont {Qiao}}, \bibinfo {author} {\bibfnamefont {F.}~\bibnamefont {Thöle}}, \bibinfo {author} {\bibfnamefont {S.~S.}\ \bibnamefont {Tsirkin}}, \bibinfo {author} {\bibfnamefont {M.}~\bibnamefont {Wierzbowska}}, \bibinfo {author} {\bibfnamefont {N.}~\bibnamefont {Marzari}}, \bibinfo {author} {\bibfnamefont {D.}~\bibnamefont {Vanderbilt}}, \bibinfo {author} {\bibfnamefont {I.}~\bibnamefont {Souza}}, \bibinfo {author} {\bibfnamefont {A.~A.}\ \bibnamefont {Mostofi}},\ and\ \bibinfo {author} {\bibfnamefont {J.~R.}\ \bibnamefont {Yates}},\ }\href
  {https://doi.org/10.1088/1361-648X/ab51ff} {\bibfield  {journal} {\bibinfo  {journal} {Journal of Physics: Condensed Matter}\ }\textbf {\bibinfo {volume} {32}},\ \bibinfo {pages} {165902} (\bibinfo {year} {2020})}\BibitemShut {NoStop}%
\bibitem [{SM(2025)}]{SM}%
  \BibitemOpen
  \href@noop {} {}\bibinfo {howpublished} {See Supplemental Material (SM) at [URL will be inserted by publisher]. Section IA contains details on comupatational setup used, Section IB contains proof that magnetic exchange is directly determined by the paths closed with non-zero hopping parameters, Section IIA contains thorough description of all contributions to the 1NN isotropic exchange in case of CrI$_{3}$ and Section IIB contains information on anisotropic magnetic exchange in case of all relevant interactions in both materials. SM includes Refs.~\cite{dudarevU, TB2J2021, moriya, Kitaev2018, amoroso2020}} (\bibinfo {year} {2025})\BibitemShut {NoStop}%
\bibitem [{\citenamefont {Dudarev}\ \emph {et~al.}(1998)\citenamefont {Dudarev}, \citenamefont {Botton}, \citenamefont {Savrasov}, \citenamefont {Humphreys},\ and\ \citenamefont {Sutton}}]{dudarevU}%
  \BibitemOpen
  \bibfield  {author} {\bibinfo {author} {\bibfnamefont {S.~L.}\ \bibnamefont {Dudarev}}, \bibinfo {author} {\bibfnamefont {G.~A.}\ \bibnamefont {Botton}}, \bibinfo {author} {\bibfnamefont {S.~Y.}\ \bibnamefont {Savrasov}}, \bibinfo {author} {\bibfnamefont {C.~J.}\ \bibnamefont {Humphreys}},\ and\ \bibinfo {author} {\bibfnamefont {A.~P.}\ \bibnamefont {Sutton}},\ }\href {https://doi.org/10.1103/PhysRevB.57.1505} {\bibfield  {journal} {\bibinfo  {journal} {Phys. Rev. B}\ }\textbf {\bibinfo {volume} {57}},\ \bibinfo {pages} {1505} (\bibinfo {year} {1998})}\BibitemShut {NoStop}%
\bibitem [{\citenamefont {Moriya}(1960)}]{moriya}%
  \BibitemOpen
  \bibfield  {author} {\bibinfo {author} {\bibfnamefont {T.}~\bibnamefont {Moriya}},\ }\href {https://doi.org/10.1103/PhysRev.120.91} {\bibfield  {journal} {\bibinfo  {journal} {Phys. Rev.}\ }\textbf {\bibinfo {volume} {120}},\ \bibinfo {pages} {91} (\bibinfo {year} {1960})}\BibitemShut {NoStop}%
\bibitem [{\citenamefont {Amoroso}\ \emph {et~al.}(2020)\citenamefont {Amoroso}, \citenamefont {Barone},\ and\ \citenamefont {Picozzi}}]{amoroso2020}%
  \BibitemOpen
  \bibfield  {author} {\bibinfo {author} {\bibfnamefont {D.}~\bibnamefont {Amoroso}}, \bibinfo {author} {\bibfnamefont {P.}~\bibnamefont {Barone}},\ and\ \bibinfo {author} {\bibfnamefont {S.}~\bibnamefont {Picozzi}},\ }\href {https://doi.org/10.1038/s41467-020-19535-w} {\bibfield  {journal} {\bibinfo  {journal} {Nature Communications}\ }\textbf {\bibinfo {volume} {11}},\ \bibinfo {pages} {5784} (\bibinfo {year} {2020})}\BibitemShut {NoStop}%
\bibitem [{\citenamefont {Fuh}\ \emph {et~al.}(2017)\citenamefont {Fuh}, \citenamefont {Chang}, \citenamefont {Hung},\ and\ \citenamefont {Jeng}}]{vi2_1}%
  \BibitemOpen
  \bibfield  {author} {\bibinfo {author} {\bibfnamefont {H.-R.}\ \bibnamefont {Fuh}}, \bibinfo {author} {\bibfnamefont {K.-W.}\ \bibnamefont {Chang}}, \bibinfo {author} {\bibfnamefont {S.-H.}\ \bibnamefont {Hung}},\ and\ \bibinfo {author} {\bibfnamefont {H.-T.}\ \bibnamefont {Jeng}},\ }\href {https://doi.org/10.1109/LMAG.2016.2621720} {\bibfield  {journal} {\bibinfo  {journal} {IEEE Magnetics Letters}\ }\textbf {\bibinfo {volume} {8}},\ \bibinfo {pages} {1} (\bibinfo {year} {2017})}\BibitemShut {NoStop}%
\bibitem [{\citenamefont {Liu}\ \emph {et~al.}(2024{\natexlab{b}})\citenamefont {Liu}, \citenamefont {Ren},\ and\ \citenamefont {Picozzi}}]{vi2_2}%
  \BibitemOpen
  \bibfield  {author} {\bibinfo {author} {\bibfnamefont {C.}~\bibnamefont {Liu}}, \bibinfo {author} {\bibfnamefont {W.}~\bibnamefont {Ren}},\ and\ \bibinfo {author} {\bibfnamefont {S.}~\bibnamefont {Picozzi}},\ }\href {https://doi.org/10.1103/PhysRevLett.132.086802} {\bibfield  {journal} {\bibinfo  {journal} {Phys. Rev. Lett.}\ }\textbf {\bibinfo {volume} {132}},\ \bibinfo {pages} {086802} (\bibinfo {year} {2024}{\natexlab{b}})}\BibitemShut {NoStop}%
\bibitem [{\citenamefont {Zhou}\ \emph {et~al.}(2022)\citenamefont {Zhou}, \citenamefont {Wang}, \citenamefont {Zhu}, \citenamefont {Liu}, \citenamefont {Hou}, \citenamefont {Guo},\ and\ \citenamefont {Zhong}}]{vi2_exp}%
  \BibitemOpen
  \bibfield  {author} {\bibinfo {author} {\bibfnamefont {X.}~\bibnamefont {Zhou}}, \bibinfo {author} {\bibfnamefont {Z.}~\bibnamefont {Wang}}, \bibinfo {author} {\bibfnamefont {H.}~\bibnamefont {Zhu}}, \bibinfo {author} {\bibfnamefont {Z.}~\bibnamefont {Liu}}, \bibinfo {author} {\bibfnamefont {Y.}~\bibnamefont {Hou}}, \bibinfo {author} {\bibfnamefont {D.}~\bibnamefont {Guo}},\ and\ \bibinfo {author} {\bibfnamefont {D.}~\bibnamefont {Zhong}},\ }\href {https://doi.org/10.1039/D2NR02367A} {\bibfield  {journal} {\bibinfo  {journal} {Nanoscale}\ }\textbf {\bibinfo {volume} {14}},\ \bibinfo {pages} {10559} (\bibinfo {year} {2022})}\BibitemShut {NoStop}%
\bibitem [{\citenamefont {Kim}\ and\ \citenamefont {Park}(2021)}]{nips_feps_67}%
  \BibitemOpen
  \bibfield  {author} {\bibinfo {author} {\bibfnamefont {T.~Y.}\ \bibnamefont {Kim}}\ and\ \bibinfo {author} {\bibfnamefont {C.-H.}\ \bibnamefont {Park}},\ }\href {https://doi.org/10.1021/acs.nanolett.1c03992} {\bibfield  {journal} {\bibinfo  {journal} {Nano Letters}\ }\textbf {\bibinfo {volume} {21}},\ \bibinfo {pages} {10114} (\bibinfo {year} {2021})},\ \Eprint {https://arxiv.org/abs/https://doi.org/10.1021/acs.nanolett.1c03992} {https://doi.org/10.1021/acs.nanolett.1c03992} \BibitemShut {NoStop}%
\bibitem [{\citenamefont {Olsen}(2021)}]{nips_feps_70}%
  \BibitemOpen
  \bibfield  {author} {\bibinfo {author} {\bibfnamefont {T.}~\bibnamefont {Olsen}},\ }\href {https://doi.org/10.1088/1361-6463/ac000e} {\bibfield  {journal} {\bibinfo  {journal} {Journal of Physics D: Applied Physics}\ }\textbf {\bibinfo {volume} {54}},\ \bibinfo {pages} {314001} (\bibinfo {year} {2021})}\BibitemShut {NoStop}%
\bibitem [{\citenamefont {Lan\ifmmode~\mbox{\c{c}}\else \c{c}\fi{}on}\ \emph {et~al.}(2018)\citenamefont {Lan\ifmmode~\mbox{\c{c}}\else \c{c}\fi{}on}, \citenamefont {Ewings}, \citenamefont {Guidi}, \citenamefont {Formisano},\ and\ \citenamefont {Wildes}}]{nips76}%
  \BibitemOpen
  \bibfield  {author} {\bibinfo {author} {\bibfnamefont {D.}~\bibnamefont {Lan\ifmmode~\mbox{\c{c}}\else \c{c}\fi{}on}}, \bibinfo {author} {\bibfnamefont {R.~A.}\ \bibnamefont {Ewings}}, \bibinfo {author} {\bibfnamefont {T.}~\bibnamefont {Guidi}}, \bibinfo {author} {\bibfnamefont {F.}~\bibnamefont {Formisano}},\ and\ \bibinfo {author} {\bibfnamefont {A.~R.}\ \bibnamefont {Wildes}},\ }\href {https://doi.org/10.1103/PhysRevB.98.134414} {\bibfield  {journal} {\bibinfo  {journal} {Phys. Rev. B}\ }\textbf {\bibinfo {volume} {98}},\ \bibinfo {pages} {134414} (\bibinfo {year} {2018})}\BibitemShut {NoStop}%
\bibitem [{\citenamefont {Wildes}\ \emph {et~al.}(2022)\citenamefont {Wildes}, \citenamefont {Stewart}, \citenamefont {Le}, \citenamefont {Ewings}, \citenamefont {Rule}, \citenamefont {Deng},\ and\ \citenamefont {Anand}}]{nips79}%
  \BibitemOpen
  \bibfield  {author} {\bibinfo {author} {\bibfnamefont {A.~R.}\ \bibnamefont {Wildes}}, \bibinfo {author} {\bibfnamefont {J.~R.}\ \bibnamefont {Stewart}}, \bibinfo {author} {\bibfnamefont {M.~D.}\ \bibnamefont {Le}}, \bibinfo {author} {\bibfnamefont {R.~A.}\ \bibnamefont {Ewings}}, \bibinfo {author} {\bibfnamefont {K.~C.}\ \bibnamefont {Rule}}, \bibinfo {author} {\bibfnamefont {G.}~\bibnamefont {Deng}},\ and\ \bibinfo {author} {\bibfnamefont {K.}~\bibnamefont {Anand}},\ }\href {https://doi.org/10.1103/PhysRevB.106.174422} {\bibfield  {journal} {\bibinfo  {journal} {Phys. Rev. B}\ }\textbf {\bibinfo {volume} {106}},\ \bibinfo {pages} {174422} (\bibinfo {year} {2022})}\BibitemShut {NoStop}%
\bibitem [{\citenamefont {Yan}\ \emph {et~al.}(2023)\citenamefont {Yan}, \citenamefont {Du}, \citenamefont {Zhang}, \citenamefont {Wan},\ and\ \citenamefont {Wang}}]{nips_feps_cops}%
  \BibitemOpen
  \bibfield  {author} {\bibinfo {author} {\bibfnamefont {S.}~\bibnamefont {Yan}}, \bibinfo {author} {\bibfnamefont {Y.}~\bibnamefont {Du}}, \bibinfo {author} {\bibfnamefont {X.}~\bibnamefont {Zhang}}, \bibinfo {author} {\bibfnamefont {X.}~\bibnamefont {Wan}},\ and\ \bibinfo {author} {\bibfnamefont {D.}~\bibnamefont {Wang}},\ }\href {https://doi.org/10.1088/1361-648X/ad06ef} {\bibfield  {journal} {\bibinfo  {journal} {Journal of Physics: Condensed Matter}\ }\textbf {\bibinfo {volume} {36}},\ \bibinfo {pages} {065502} (\bibinfo {year} {2023})}\BibitemShut {NoStop}%
\bibitem [{\citenamefont {Wildes}\ \emph {et~al.}(2012)\citenamefont {Wildes}, \citenamefont {Rule}, \citenamefont {Bewley}, \citenamefont {Enderle},\ and\ \citenamefont {Hicks}}]{feps77}%
  \BibitemOpen
  \bibfield  {author} {\bibinfo {author} {\bibfnamefont {A.~R.}\ \bibnamefont {Wildes}}, \bibinfo {author} {\bibfnamefont {K.~C.}\ \bibnamefont {Rule}}, \bibinfo {author} {\bibfnamefont {R.~I.}\ \bibnamefont {Bewley}}, \bibinfo {author} {\bibfnamefont {M.}~\bibnamefont {Enderle}},\ and\ \bibinfo {author} {\bibfnamefont {T.~J.}\ \bibnamefont {Hicks}},\ }\href {https://doi.org/10.1088/0953-8984/24/41/416004} {\bibfield  {journal} {\bibinfo  {journal} {Journal of Physics: Condensed Matter}\ }\textbf {\bibinfo {volume} {24}},\ \bibinfo {pages} {416004} (\bibinfo {year} {2012})}\BibitemShut {NoStop}%
\bibitem [{\citenamefont {Lan\ifmmode~\mbox{\c{c}}\else \c{c}\fi{}on}\ \emph {et~al.}(2016)\citenamefont {Lan\ifmmode~\mbox{\c{c}}\else \c{c}\fi{}on}, \citenamefont {Walker}, \citenamefont {Ressouche}, \citenamefont {Ouladdiaf}, \citenamefont {Rule}, \citenamefont {McIntyre}, \citenamefont {Hicks}, \citenamefont {R\o{}nnow},\ and\ \citenamefont {Wildes}}]{feps78}%
  \BibitemOpen
  \bibfield  {author} {\bibinfo {author} {\bibfnamefont {D.}~\bibnamefont {Lan\ifmmode~\mbox{\c{c}}\else \c{c}\fi{}on}}, \bibinfo {author} {\bibfnamefont {H.~C.}\ \bibnamefont {Walker}}, \bibinfo {author} {\bibfnamefont {E.}~\bibnamefont {Ressouche}}, \bibinfo {author} {\bibfnamefont {B.}~\bibnamefont {Ouladdiaf}}, \bibinfo {author} {\bibfnamefont {K.~C.}\ \bibnamefont {Rule}}, \bibinfo {author} {\bibfnamefont {G.~J.}\ \bibnamefont {McIntyre}}, \bibinfo {author} {\bibfnamefont {T.~J.}\ \bibnamefont {Hicks}}, \bibinfo {author} {\bibfnamefont {H.~M.}\ \bibnamefont {R\o{}nnow}},\ and\ \bibinfo {author} {\bibfnamefont {A.~R.}\ \bibnamefont {Wildes}},\ }\href {https://doi.org/10.1103/PhysRevB.94.214407} {\bibfield  {journal} {\bibinfo  {journal} {Phys. Rev. B}\ }\textbf {\bibinfo {volume} {94}},\ \bibinfo {pages} {214407} (\bibinfo {year} {2016})}\BibitemShut {NoStop}%
\bibitem [{\citenamefont {Gu}\ \emph {et~al.}(2019)\citenamefont {Gu}, \citenamefont {Zhang}, \citenamefont {Le}, \citenamefont {Li}, \citenamefont {Xiang},\ and\ \citenamefont {Hu}}]{cops_68}%
  \BibitemOpen
  \bibfield  {author} {\bibinfo {author} {\bibfnamefont {Y.}~\bibnamefont {Gu}}, \bibinfo {author} {\bibfnamefont {Q.}~\bibnamefont {Zhang}}, \bibinfo {author} {\bibfnamefont {C.}~\bibnamefont {Le}}, \bibinfo {author} {\bibfnamefont {Y.}~\bibnamefont {Li}}, \bibinfo {author} {\bibfnamefont {T.}~\bibnamefont {Xiang}},\ and\ \bibinfo {author} {\bibfnamefont {J.}~\bibnamefont {Hu}},\ }\href {https://doi.org/10.1103/PhysRevB.100.165405} {\bibfield  {journal} {\bibinfo  {journal} {Phys. Rev. B}\ }\textbf {\bibinfo {volume} {100}},\ \bibinfo {pages} {165405} (\bibinfo {year} {2019})}\BibitemShut {NoStop}%
\bibitem [{\citenamefont {Kim}\ \emph {et~al.}(2020)\citenamefont {Kim}, \citenamefont {Jeong}, \citenamefont {Park}, \citenamefont {Masuda}, \citenamefont {Asai}, \citenamefont {Itoh}, \citenamefont {Kim}, \citenamefont {Wildes},\ and\ \citenamefont {Park}}]{cops_75}%
  \BibitemOpen
  \bibfield  {author} {\bibinfo {author} {\bibfnamefont {C.}~\bibnamefont {Kim}}, \bibinfo {author} {\bibfnamefont {J.}~\bibnamefont {Jeong}}, \bibinfo {author} {\bibfnamefont {P.}~\bibnamefont {Park}}, \bibinfo {author} {\bibfnamefont {T.}~\bibnamefont {Masuda}}, \bibinfo {author} {\bibfnamefont {S.}~\bibnamefont {Asai}}, \bibinfo {author} {\bibfnamefont {S.}~\bibnamefont {Itoh}}, \bibinfo {author} {\bibfnamefont {H.-S.}\ \bibnamefont {Kim}}, \bibinfo {author} {\bibfnamefont {A.}~\bibnamefont {Wildes}},\ and\ \bibinfo {author} {\bibfnamefont {J.-G.}\ \bibnamefont {Park}},\ }\href {https://doi.org/10.1103/PhysRevB.102.184429} {\bibfield  {journal} {\bibinfo  {journal} {Phys. Rev. B}\ }\textbf {\bibinfo {volume} {102}},\ \bibinfo {pages} {184429} (\bibinfo {year} {2020})}\BibitemShut {NoStop}%
\end{thebibliography}%

\end{document}